\documentclass[review]{elsarticle}

\usepackage{hyperref,bm}
\usepackage{amsmath}

\journal{Journal of \LaTeX\ Templates}

\begin{document}

\begin{frontmatter}

\title{An Hourglass Control Algorithm for Lagrangian Smooth Particle Hydrodynamics}
\author[mymainaddress]{Georg C. Ganzenm\"uller \corref{mycorrespondingauthor}}
\cortext[mycorrespondingauthor]{Corresponding author}
\ead{georg.ganzenmueller@emi.fraunhofer.de}

\address[mymainaddress]{Fraunhofer Ernst-Mach Institute for High-Speed Dynamics, EMI, Eckerstr. 4,
79104 Freiburg, Germany}

\begin{abstract}
This paper presents a stabilization scheme which addresses the rank-deficiency problem in meshless
collocation methods for solid mechanics. Specifically, Smooth-Particle Hydrodynamics (SPH) in the
Total Lagrangian formalism is considered. This method is rank-deficient in the sense that the SPH
approximation of the deformation gradient is not unique with respect to the positions of the
integration points. The non-uniqueness can result in the formation of zero-energy modes. If
undetected, these modes can grow and completely dominate the solution. Here, an algorithm is
introduced, which effectively suppresses these modes in a fashion similar to hour-glass control
mechanisms in Finite-Element methods.  Simulations utilizing this control algorithm result exhibit
much improved stability, accuracy, and error convergence properties. In contrast to an alternative
method which eliminates zero-energy modes, namely the use of additional integration points, the here
presented algorithm is easy to implement and computationally very efficient.
\end{abstract}


\end{frontmatter}



\section{Introduction}

Smooth-Particle Hydrodynamics (SPH) was first developed in the seventies by Lucy \cite{Lucy:1977/a} and Gingold and
Monaghan \cite{Gingold:1977a} in the astrophysical context as a means to simulate the formation of
stars. The principle of SPH is to approximate a continuous field using a discrete set of kernel
functions which are centered about so-called particles, where the physical properties of the system,
e.g. mass, internal energy, or velocity, are located. When the SPH approximation is applied to fluid
flow or solid body deformation, solutions to the underlying set of partial differential equations
are obtained in terms of simple algebraic equations. As no auxiliary computational grid is used to
construct the, SPH is termed a mesh-free method. The absence of a mesh implies that arbitrarily
large deformations and instability phenomena such as fracture can be handled with ease when compared
to mesh-based techniques such as the Finite Element method (FEM). FEM requires geometrically
well-defined mesh cells and certain assumptions regarding the smoothness of the field within these cells.
Because both of these requirements are typically violated in the simulation of important engineering
applications such as impact, explosion, or machine cutting, a continued interest in the development
of meshfree methods prevails.

The application of SPH to fluid problems with free boundaries has been very successful. In
particular, solutions to large-scale gas dynamic problems have been obtained in astrophysics
\cite{Springel:2010/a}, and fluid-structure interaction in civil engineering applications, such as
the impact of a water wave on coastal structures, has been treated with success
\cite{Gomez-Gesteira:2010/a}.  However, for solid body deformations, the situation is not equally
satisfying. In 1991, Libersky \cite{Libersky:1991/a} was the first to simulate a body with material
strength using SPH. Strong numerical instability issues appeared which prevented SPH from becoming a
serious competitor to mesh-based methods for solid continua. Different sources of instability were
identified by early works: Swegle \cite{Swegle:1995/a} noted that the interaction of the second
derivative of the kernel and the tensile stress resulted in nonphysical clumping of particles, which
he termed \textit{tensile instability}. Dyka \textit{et al} \cite{Dyka:1997/a} observed that the
nodal integration approach inherent to SPH incurs instabilities. In essence, the number of
integration points is too small such that the solution to the underlying equilibrium equation
becomes non-unique due to rank deficiency.  They proposed to eliminate the rank-deficiency by
introducing additional integration points at other locations that the particles themselves and noted
that this kind of instability is also observed in FEM, when elements with a reduced number of
integration points are used. In FEM, the instability emerging from this rank-deficiency problem is
termed hour-glassing.  Rank deficiency occurs regardless of the state of stress and in addition to
the tensile instability.  A number of different schemes were devised to increase the stability of
SPH. Artificial viscosity and Riemann-type solvers increase numerical stability by dissipating
high-frequency modes. Conservative smoothing \textit{et al} \cite{Hicks:1997/a} and the XSPH time
integration scheme \cite{Monaghan:1989/a} are dispersive rather than dissipative but also work by
removing high-frequency modes.  Randles \textit{et al} \cite{Randles:1996a} elaborated on the idea of
introducing additional stress points to remove the rank-deficiency problem. The clumping problem
associated with tensile instability was addressed by Gray \textit{et al.} \cite{Gray:2001/a} by 
adding repulsive forces between SPH particles if the principal stresses are tensile. However, none of these
approaches turned SPH into a simulation method that is generally stable for a broad range
of applications.

A turning point was achieved by Belytschko \textit{et al.} \cite{Belytschko:2000/a}, who showed that
the Eulerian character of the kernel function (other particles pass through a particle's kernel
domain as the simulation proceeds) in combination with the Lagrangian character of the moving SPH
particles (they move in a fixed frame of reference) is the cause of the tensile instability. They
proposed a Lagrangian formulation where the kernel approximation is performed in the initial,
undeformed reference coordinates of the material. In this Lagrangian formulation, the tensile
instability is absent, however, other instabilities due to rank-deficiency caused by the collocation
method remain.  Belytschko \textit{et al} also showed that the remaining instabilities can be
removed by the addition of stress points, but the location of the stress points needs to be
carefully chosen.  The invention of Lagrangian SPH has prompted a revived interest in this
particular meshless method, resulting in several studies which confirm its enhanced stability
\cite{Bonet:2001/a, Bonet:2002/a, Rabczuk:2004a, Vignjevic:2006a, Xiao:2005/a}.  However, the idea of using additional
integration points appears not to have found widespread usage in the SPH community.  Possible causes
might be related to the increased computational effort required for evaluating the stresses and a
lack of information as to where stress points should be placed if irregular particle positions are
employed for the reference configuration. 

In this work, it is demonstrated how instabilities caused by the rank-deficiency can be directly controlled.
The inspiration for taking such an approach to stabilize the solution originates from ideas
developed for FEM. There, elements with a reduced number of integration points are routinely employed because they
are computationally very effective and avoid the shear locking problems of fully integrated
elements. Such reduced-integrated elements are susceptible to so-called hourglass modes, which are
zero-energy modes in the sense that the element deforms without an associated increase of the
elastic energy. These modes cannot be detected if a reduced number of integration points is used, and
can therefore be populated with arbitrary amounts of kinetic energy, such that the solution is
entirely dominated by theses modes. A common approach to suppress the hour-glassing modes is to
identify them as the non-linear part of the velocity field and penalize them by appropriate means.
It is difficult in general to seek analogies between the SPH collocation method and FEM. However, in
the case of the so-called mean (or constant) stress element \cite{Flanagan:1981/a}, which uses only one
integration point to represent the average stress state within the entire element, there exists a clear
analogy to the weighted average over the neighboring particles that is obtained in SPH. It is this
analogy that will be exploited in order to develop a zero-energy mode suppression algorithm for SPH.

The remainder of this article is organized as follows.  In the next section, a brief review of the
Lagrangian SPH formalism is given. This is followed by the details as to how an SPH analogue of the
non-linear part of the deformation field can be used to obtain an algorithm which effectively
suppresses zero-energy modes.  The usefulness of the stabilization algorithm is subsequently
demonstrated with a number of large strain deformation examples, that are difficult, if not
impossible, to obtain using Lagrangian SPH without additional stress points. Finally, the
implications of this particular type of stabilization technique are discussed, and an outlook is
given regarding possible improvements.


\section{Lagrangian SPH}
Smooth Particle Hydrodynamics was originally devised as a Lagrangian particle method with the the
smoothing kernel travelling with the particle, thus redefining the interaction neighbourhood for
every new position the particle attains. In this sense, the kernel of the original SPH formulation
has Eulerian character, as other particles move through the kernel domain. The tensile instability
\cite{Swegle:1995/a} encountered in SPH, where particles clump together under negative pressure
conditions, has been found to be caused by the Eulerian kernel functions \cite{Belytschko:2000/a}.
Consequently, formulations were developed \cite{Belytschko:2000/a, Bonet:2002/a, Rabczuk:2004a,
Vignjevic:2006a}, which use define the kernel functions in a fixed reference configuration.
Typically, the initial, undeformed configuration of particles is taken as the reference
configuration. This particular flavor of SPH is referred to as \textit{Lagrangian SPH}
\cite{Belytschko:2000/a, Bonet:2002/a}, or \textit{Total Lagrangian SPH} \cite{Vignjevic:2006a}. In
the remainder of this article, the latter is used, as it reminds more clearly of the reference
configuration aspect.  In what follows, SPH and the associated nomenclature is
briefly explained with the limited scope of obtaining a set of SPH expressions suitable to describe
the deformation of a solid body. For more detailed derivations, the reader is referred to the
works cited above.

\subsection{The Total Lagrangian formulation}
In the Total Lagrangian formulation, conservation equations and constitutive equations are expressed
in terms of the reference coordinates $\bm{X}$, which are taken to be the coordinates of the initial,
undeformed reference configuration. A mapping $\phi$ between the current coordinates, and the reference
coordinates describes the body motion at time $t$:
\begin{equation}
	\bm{x} = \phi(\bm{X}, t),
\end{equation}
Here, $\bm{x}$ are the current, deformed coordinates and $\bm{X}$ the reference (Lagrangian) coordinates. The
displacement $\bm{u}$ is given by
\begin{equation}
	\bm{u} =\bm{x} - \bm{X},
\end{equation}
Note that bold mathematical symbols like the preceding ones denote vectors, while the same
mathematical symbol in non-bold font refers to its Euclidean norm, e.g. $x = |\bm{x}|$.
The conservation equations for mass, impulse, and energy in the total Lagrangian formulation are given by
\begin{eqnarray}
	\rho J & = & \rho_0\\
	\ddot{u} & = & \frac{1}{\rho_0} \bm{\nabla}_0 \cdot \bm{P}^{T} \label{eqn:divergence}\\
	\dot{e} & = & \frac{1}{\rho_0} \dot{\bm{F}}:\bm{P},
\end{eqnarray}
where $\rho$ is the mass density, $\bm{P}$ is the first Piola-Kirchhoff stress
tensor, $e$ is the internal energy, and $\bm{\nabla}$ is the gradient or
divergence operator. The subscript 0 indicates that a quantity is evaluated in
the reference configuration, while the absence of this subscript means that the
current configuration is to be used.  $J$ is the determinant of the deformation
gradient $\bm{F}$, 
\begin{equation} \label{eqn:defgrad}
	\bm{F} = \frac{\mathrm{d}\bm{x}} {\mathrm{d}\bm{X}} = \frac{\mathrm{d}\bm{u}}
{\mathrm{d}\bm{X}} + \bm{I},
\end{equation}
which can be interpreted as the transformation matrix that describes the rotation and stretch of a
line element from the reference configuration to the current configuration.


\subsection {The SPH Approximation}
The SPH approximation for a scalar function $f$ in terms of the reference coordinates can be
written as
\begin{equation} \label{eqn:sph_function_approximation}
	f(\bm{X_i}) = \sum \limits_{j \in \mathcal{S}} V_j^0 f(\bm{X_j}) W_i \left( \bm{X}_{ij} \right)
\end{equation}
The sum extends over all particles within the range of a scalar weight function $W_i$, which is
centered at position $\bm{X_i}$ and depends only on the distance vector between coordinates
$\bm{X}_i$ and $\bm{X}_j$. Here, exclusively radially symmetric kernels are considered, i.e., $W_i
\left( \bm{X}_{ij} \right) = W_i \left( X_{ij} \right)$ which depend only on the scalar distance
between particles $i$ and $j$.  $V^0$ is the volume associated with a particle in the reference
configuration. The weight function is chosen to have compact support, i.e., it includes only
neighbors within a certain radial distance. This domain of influence is denoted $\mathcal{S}$.

The SPH approximation of a derivative of $f$ is obtained by operating directly with the gradient
operator on the kernel functions,
\begin{equation} \label{eqn:sph_derivative_approximation}
	\bm{\nabla}_0 f(\bm{X_i}) = \sum \limits_{j \in \mathcal{S}} V_j^0 f(\bm{X_j}) \bm{\nabla} W_i \left(
X_{ij} \right),
\end{equation}
where the gradient of the kernel function is defined as follows:
\begin{equation}
	\bm{\nabla} W_i(X_{ij}) = \left ( \frac{\mathrm{d}W(X_{ij})}{\mathrm{d}X_{ij}} \right ) \frac{\bm{X}_j - \bm{X}_i}
{X_{ij}}
\end{equation}
It is of fundamental interest to characterize numerical approximation methods
in terms of the order of completeness, i.e., the order of a polynomial that can
be exactly approximated by the method. For solving the conservation equations
with its differential operators, at least first-order completeness is required.
In the case of the SPH approximation, the conditions for zero$^{th}$- and
first-order completeness are stated as follows:
\begin{eqnarray}
	\sum \limits_{j \in \mathcal{S}} V_j^0 W_i \left( X_{ij} \right) = 1
\label{eqn:zeroth_order_completeness}\\
	\sum \limits_{j \in \mathcal{S}} V_j^0 \bm{\nabla} W_i \left( X_{ij} \right) = 0 \label{eqn:first_order_completeness}
\end{eqnarray}
The basic SPH approach given by equations (\ref{eqn:sph_function_approximation}) and
(\ref{eqn:sph_derivative_approximation}) fulfills neither of these completeness conditions.  An
\textit{ad-hoc} improvement by Monaghan \cite{Monaghan:1988/a} consists of adding
eqn.~(\ref{eqn:first_order_completeness}) to eqn.~(\ref{eqn:sph_derivative_approximation}), such
that a \textit{symmetrized} approximation for the derivative of a function is obtained,
\begin{equation} \label{eqn:sph_derivative_approximation_symmetrized}
	\bm{\nabla}_0 f(\bm{X_i}) = \sum \limits_{j \in \mathcal{S}} V_j^0 \left( f(\bm{X_j}) - f(\bm{X_i})
\right ) \bm{\nabla} W_i \left( X_{ij} \right)
\end{equation}
The symmetrization does not result in first-order completeness, however, it yields zeroth-order
completeness for the derivatives of a function, even in the case of irregular particle arrangements
\cite{Rabczuk:2004a}.

\subsection {First-Order Completeness}
In order to fulfill first-order completeness, the SPH approximation has to reproduce the constant
gradient of a linear field. A number of correction techniques \cite{Randles:1996a, Bonet:1999a,
Vignjevic:2000a} exploit this condition as the basis for correcting the gradient of the SPH weight
function,
\begin{equation} \label{eqn:first_order_completeness_vector}
	\sum \limits_{j \in \mathcal{S}} V_j^0 (\bm{X}_j - \bm{X_i}) \otimes \bm{\nabla} W_i(X_{ij}) \overset{!}{=} \bm{I},
\end{equation}
where $\bm{I}$ is the diagonal unit matrix. Based on this expression, a corrected kernel gradient can be
defined:
\begin{equation} \label{eqn:correct_SPH_gradient}
	\tilde{\bm{\nabla}} W_i(X_{ij}) = \bm{L}_i^{-1} \bm{\nabla} W_i(X_{ij}),
\end{equation}
which uses the correction matrix $\bm{L}$, given by:
\begin{equation} \label{eqn:SPH_correction_matrix}
	\bm{L}_i = \sum \limits_{j \in \mathcal{S}} V_j^0 \bm{\nabla} W_i(X_{ij}) \otimes (\bm{X}_j -
	\bm{X_i}).
\end{equation}
By construction, the corrected kernel gradient now satisfy eqn.~(\ref{eqn:first_order_completeness_vector}),
\begin{equation} \label{eqn:first_order_completeness_vector)}
	\sum \limits_{j \in \mathcal{S}} V_j^0 (\bm{X}_j - \bm{X_i}) \otimes \bm{L}_i^{-1} \bm{\nabla}  W_i(X_{ij})
= \bm{I},
\end{equation}
resulting in first-order completeness.

\subsection{Total-Lagrangian SPH expressions for Solid Mechanics}
For calculating the internal forces of a solid body subject to deformation, expressions are required
for (i) the deformation gradient, (ii) a constitutive equation which provides a stress tensor as
function of the deformation gradient, and (iii) an expression for transforming the stresses into
forces acting on the nodes which serve as the discrete representation of the body.

\subsubsection{the Deformation Gradient}
The deformation gradient is obtained by calculating the derivative of the displacement field, i.e.
by using the symmetrized SPH derivative approximation,
eqn.~(\ref{eqn:sph_derivative_approximation_symmetrized}), for eqn.~(\ref{eqn:defgrad}):
\begin{equation} \label{eqn:SPH_defgrad}
	\bm{F}_i = \sum \limits_{j \in \mathcal{S}} V_j^0 (\bm{u}_j - \bm{u_i}) \otimes \bm{L}_i^{-1} \bm{\nabla}
W_i(X_{ij}) + \bm{I}.
\end{equation}
Note that in the above equation, the corrected kernel gradients have been introduced.

\subsubsection{Constitutive Models}
The constitutive model is independent of the numerical discretization and therefore no essential
part of the SPH method. However, some important relations are quoted for clarity and the
reader is referred to the excellent textbooks by Bonet and Wood \cite{Bonet:2008/a} or Belytschko \textit{et al}
\cite{Belytschko:2014/a}. In the Total-Lagrangian form, the natural strain measure is the
Green-Lagrange strain,
\begin{equation} \label{eqn:Green-Lagrange}
	\bm{E} = \frac{1}{2} \left( \bm{C}  - \bm{I}\right),
\end{equation}
where $\bm{C}=\bm{F}^T \bm{F}$ is the right Cauchy-Green deformation tensor which describes changes
of line elements in the reference configuration. The proper stress measure is the Second
Piola-Kirchhoff stress $\bm{S}$, as it is work-conjugate to the Green-Lagrange strain. It is noted that
infinitesimal strain constitutive models, which are typically expressed using the Cauchy stress and an
infinitesimal strain measure, may be used in the Total Lagrangian Formulation, provided that the
Cauchy stress is replaced by $\bm{S}$ and that the infinitesimal strain is replaced by the
Green-Lagrange strain. For example, the linear elasticity model is given in terms of the Lam\'e
parameters $\lambda$ and $\mu$ as
\begin{equation}
	\bm{S} = \lambda \mathrm{Tr}\{\bm{E}\} + 2 \mu \bm{E},
\end{equation}
where $\mathrm{Tr}\{\bm{E}\}=E_{ii}/3$ denotes the trace of $\bm{E}$.  The Second Piola-Kirchhoff
stress tensor applies to the reference configuration, however, nodal forces are applied in the
current configuration.  Therefore, a stress measure is required which links the stress in the
reference configuration to the current configuration. This stress measure is the First
Piola-Kirchhoff stress, given by
\begin{equation}
	\bm{P} = \bm{F} \bm{S}.
\end{equation}

\subsubsection{Nodal Forces}
Nodal forces are obtained from an SPH approximation of the stress divergence,
eqn.~(\ref{eqn:divergence}). Several different approximations can be obtained \cite{Gingold:1982a},
depending on how the discretization is performed. The most frequently used expression, which is
variationally consistent in the sense that it minimizes elastic energy \cite{Bonet:1999a}, is the
following,
\begin{equation} \label{eqn:SPH_force_uncorrected}
	\bm{f}_i = \sum \limits_{j \in \mathcal{S}} V_i^0 V_j^0 \left ( \bm{P}_j + \bm{P}_i \right )
	\bm{\nabla} W_i(X_{ij}),
\end{equation}
where the stress tensors are added to each other rather than subtracted from each other.
For a radially symmetric kernel which depends only on distance, the
anti-symmetry property $\bm{\nabla} W_i(X_{ij}) = - \bm{\nabla} W_j(X_{ji})$
holds. Therefore, the above force expression will conserve linear momentum
exactly, as $\bm{f}_{ij} = - \bm{f}_{ji}$. The anti-symmetry property of
the kernel gradient is used to rewrite the force expression as follows:
\begin{eqnarray} \label{eqn:SPH_forces}
	\bm{f}_{i} & = & \sum \limits_{j \in \mathcal{S}} V_i^0 V_j^0 \left ( \bm{P}_i \bm{\nabla} W_i(X_{ij}) + \bm{P}_j \bm{\nabla} W_i(X_{ij}) \right ) \\
                 & = & \sum \limits_{j \in \mathcal{S}} V_i^0 V_j^0 \left ( \bm{P}_i \bm{\nabla} W_i(X_{ij}) - \bm{P}_j \bm{\nabla} W_j(X_{ji}) \right ).
\end{eqnarray}
Replacing the uncorrected kernel gradients with the corrected gradients (c.f.
eqn.~(\ref{eqn:correct_SPH_gradient}), the following expression is obtained:
\begin{equation} \label{eqn:SPH_stress_forces}
	\bm{f}_{i} = \sum \limits_{j \in \mathcal{S}} V_i^0 V_j^0 \left ( \bm{P}_i \bm{L}_i^{-1} \bm{\nabla} W_i(X_{ij}) - \bm{P}_j
\bm{L}_j^{-1} \bm{\nabla} W_j(X_{ji}) \right )
\end{equation}
This first-order corrected force expression also conserves linear momentum due to its anti-symmetry with respect
to interchange of the particle indices $i$ and $j$, i.e., $\bm{f}_{ij} = - \bm{f}_{ji}$.  The here
constructed anti-symmetric force expression is usually not seen in the literature. In contrast, it
seems to be customary \cite{Randles:1996a, Bonet:1999a, Vignjevic:2000a} to directly insert the
corrected kernel gradient into eqn.~(\ref{eqn:SPH_force_uncorrected}), which destroys the local
conservation of linear momentum.
This section is summarized by noting that all expressions for Total-Lagrangian SPH have now been
defined. The next section will introduce an SPH analogue of the hour-glassing control mechanism used
in FEM.


\section{A suppression algorithm for zero-energy modes}
Having recognized that zero-energy modes somewhat similar to hourglass modes in Finite Element
methods may exist in SPH, approach to suppress these modes can now be derived. This approach is
inspired by Flanagan and Belytschko \cite{Flanagan:1981/a}, who described hour-glassing correction
techniques for underintegrated finite elements with a linear basis. Such elements employ only a
single integration point for computing a mean deformation gradient, resulting in a mean stress and
strain value over the entire element.  Flanagan and Belytschko recognized that the mean
stress-strain description can only represent a fully linear velocity field. This implies, that all
nodal displacements have to be described exactly by the deformation gradient, which is itself a
linear operator. Nodal displacements or velocities incompatible with the linear field are identified
as the hourglass modes and corrected by suitable means.

SPH bears some similarity to the mean stress-strain description of a one-integration point finite
element: the kernel approximation leads to a smeared-out description of field variables defined at
the SPH particle's center. While the entire field over the simulation domain may vary, a single
particle assumes a constant (or mean) field within its neighborhood. This property of SPH is
directly related to the nodal integration approach, which is a piecewise constant integration
technique.

In view of this similarity, the displacement field in the neighborhood of a particle is required to
be linear. Therefore, it has to be exactly described by the deformation gradient, and the SPH
hourglass modes are identified as that part of the displacement field, which is not described by the
deformation gradient.  In practice, linear displacements are obtained by operating with the
deformation gradient on line elements of the reference configuration. The linear displacements are
compared with actual displacements in the current configuration. An error vector is defined as the
difference of these displacements, and a penalty force is applied which minimizes the error vector.

\subsection{The correction force}
\begin{figure}[!ht]
\begin{center}
\includegraphics[width=0.75\textwidth]{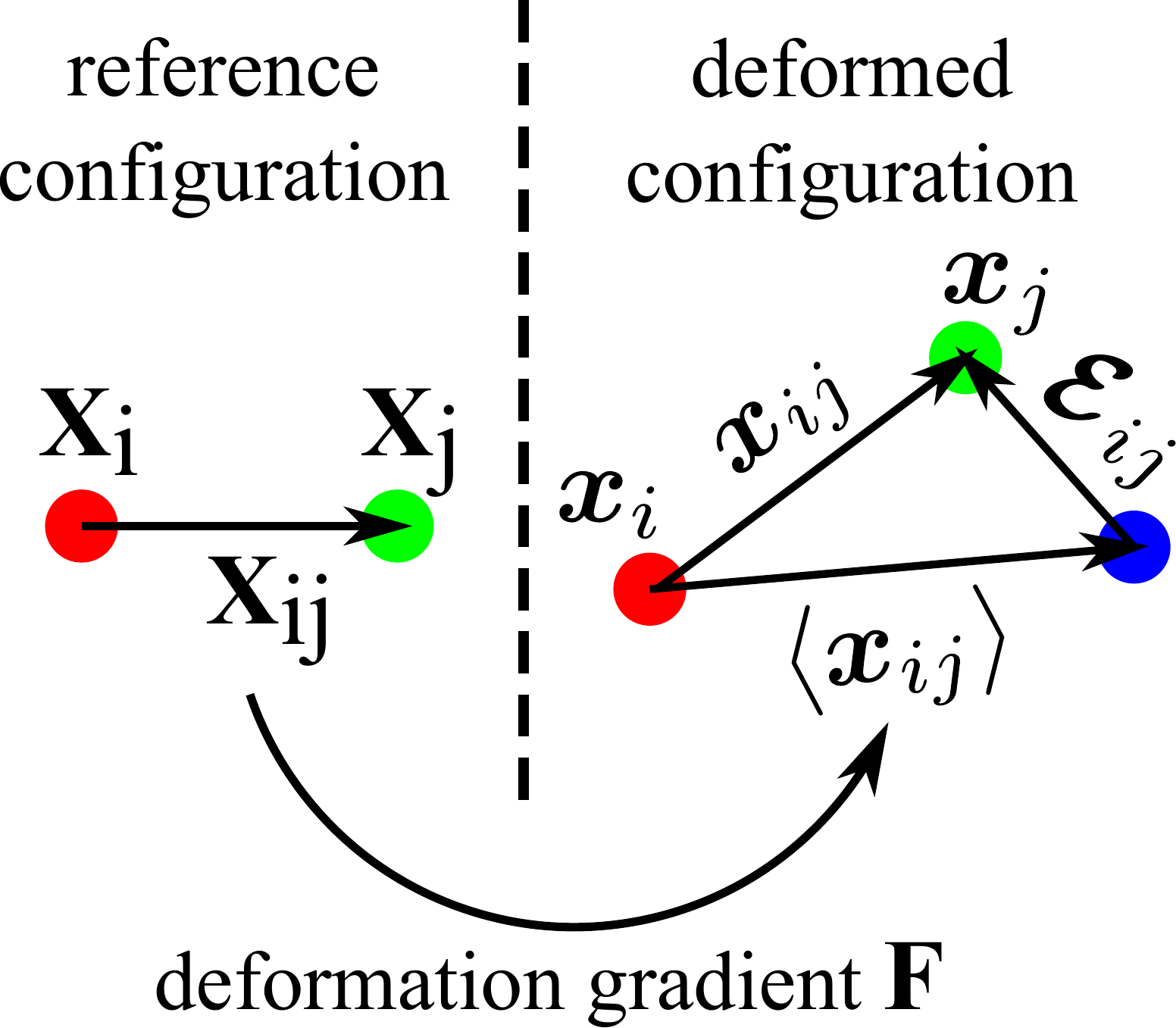}
\end{center}
\caption{ Illustration of the SPH stabilization scheme
presented here. For any pair of particles $(i,j)$, the deformed image of the reference configuration distance vector
$\left< \bm{x}_{ij} \right> = \bm{F} \bm{X}_{ij}$ is compared against the actual distance vector of
the deformed coordinates, $\bm{x}_{ij} = \bm{x}_{j} - \bm{x}_{i}$. The difference between these
vectors defines the error vector,  $\bm{\mathcal{E}}_{ij} = \left< \bm{x}_{ij} \right> -
\bm{x}_{ij}$, which is minimized by a penalty force.}
\label{fig:stabilization_scheme}
\end{figure}

To illustrate the idea of the SPH equivalent of an hourglass-control algorithm, consider the
distance vector $\bm{X}_{ij}$ between a central node $i$ and a neighboring particle $j$ in the
reference configuration.
Applying the SPH approximation of the deformation gradient to this vector
results in the \textit{approximated relative separation} $\left< \bm{x}_{ij} \right>^i$,
\begin{equation}
	\left< \bm{x}_{ij} \right>^i = \bm{F}_i \bm{X}_{ij}.
\end{equation}
Note that the superscript $i$ denotes the application of $\bm{F}_i$ to $\bm{X}_{ij}$.
Ideally, if the deformation gradient is exact, and there exists a unique mapping between all
particles' positions within the neighborhood of the central node and the deformation gradient, the
approximated relative separation will coincide with the actual relative separation of the
particles $i$ and $j$ in the deformed configuration. This situation is, however, typically not the
case due the presence of integration errors.
The difference of the approximated and actual relative separations defines the error vector, 
\begin{equation}
	\bm{\mathcal{E}}_{ij}^i = \left< {\bm{x}}_{ij} \right>^i - \bm{x}_{ij}.
\end{equation}
Figure \ref{fig:stabilization_scheme} gives a graphical representation of the error vector.
To stabilize the system and suppress the hourglass modes, the magnitude of the error vector needs to
be minimized. In this work, we formulate a correction force which is proportional to the magnitude of
the error vector, i.e., we add stiffness to the system to counteract the hourglass modes.
It is advantageous to formulate the correction force in terms of pairwise particle forces, as this
approach conserves linear and angular momentum if the force is anti-symmetric with respect to
interchange of the particle indices and co-linear with the particle separation vector. To this end,
we measure that part of the error vector which points along the direction of the current particle separation
vector:
\begin{equation}
	\delta_{ij}^i = \frac{ \bm{\mathcal{E}}_{ij}^i \cdot \bm{x}_{ij} } {x_{ij}}.
\end{equation}
$\delta_{ij}^i$ is a scalar with dimensions of length that indicates the change of distance which the
particle separation should attain in order to minimize the error vector. In what follows, a force
that is linear in $\delta_{ij}^i$ is derived, and its stiffness is expressed using the Young's
modulus of the system.

Consider the potential energy density $\hat{\phi}$ (energy per unit volume) of a Hookean material subjected to uniaxial strain, i.e., a linear spring,
\begin{equation}
	\label{eqn:hookean_spring}
	\hat{\phi} = \frac{1}{2} E \varepsilon^2,
\end{equation}
where $E$ is the Young's modulus and $\varepsilon=(L-L_0)/L_0$ is the uniaxial strain, with $L$ and
$L_0$ being the current and rest length of the spring. The force density of this spring is
\begin{equation}
	\label{eqn:hookean_spring_force}
	\hat{f} = -\frac{\mathrm{d} \hat{\phi}} {\mathrm{d} L} =  - \frac{E}{L_0} \varepsilon.
\end{equation}
In the context of the hourglass correction force, deviations of the length of the spring away from
its rest length are due to a non-zero $\delta_{ij}^i$, and the rest length of the spring is the
particle separation in the reference configuration. Thus, the hourglass correction force per unit
volume between particles $i$ and $j$, and obtained using the deformation gradient of particle $i$
is:
\begin{equation}
	\label{eqn:hookean_spring_force2}
	\hat{\bm{f}}_{ij}^i = - \frac{E}{X_{ij}} \frac{\delta_{ij}^i }{X_{ij}} \frac{\bm{x}_{ij}}{x_{ij}}.
\end{equation}
The above hourglass correction force can be improved in two aspects.
(i) To be consistent with the SPH picture, where particle interactions are weighted
according to volume and distance, a normalized smoothing kernel is introduced. (ii) The hourglass
force is unsymmetrical with respect interchange of particle indices. Therefore, explicit
symmetrization via the arithmetic mean is used. Thus, the expression for the smoothed and
symmetrized correction force density between particles $i$ and $j$ reads:
\begin{equation}
	\label{eqn:hourglass_force_ij}
	\hat{\bm{f}}_{ij}^{HG} = -\frac{1}{2} \, \alpha \, W_{ij} \, \left[\hat{\bm{f}}_{ij}^i
+ \hat{\bm{f}}_{ji}^j \right],
\end{equation}
where $\alpha$ is a dimensionless coefficient that controls the amplitude of hourglass
correction, and $W_{ij}=W_{ij}\left( X_{ij} \right)$ is a smoothing kernel which depends on the
distance in the reference configuration. Finally, the expression for the total hourglass correction force on
particle $i$ due to all neighbors j is,
\begin{eqnarray}
	\bm{f}_{i}^{HG} & = & V_i \sum \limits_{j \in \mathcal{S}} V_j W_{ij} \hat{\bm{f}}_{ij}^{HG} \nonumber \\
                        & = & \sum \limits_{j \in \mathcal{S}} -\frac{1}{2} \, \alpha\, \frac{V_i V_j W_{ij}}
{X_{ij}^2} \, \left[E_i \delta_{ij}^i + E_j \delta_{ji}^j \right] \, \frac{\bm{x}_{ij}}{x_{ij}}.
\label{eqn:hg_force}
\end{eqnarray}
This expression preserves both linear and angular momentum as the term within the sum is anti-symmetric and
collinear with the particle separation vector.

\subsection{Similarity to Laplacian stabilization}
The above introduced SPH stabilization approach is derived using arguments which originate in
hourglass control schemes for Lagrangian Finite Element Methods. It is noteworthy, however, that an
analogy to a Laplacian stabilization scheme can also be derived, which is used with Eulerian
Computational Fluid Dynamic (CFD) methods. The analogy will be derived for the one-dimensional case.

Following a classic paper by Jameson an Mavripilis \cite{Jameson:1985/a}, the flux of conserved
quantities between adjacent cells is described by the following  evolution equation,
\begin{equation}
\frac{\mathrm{d}}{\mathrm{d}t} S_i w_i = - Q(w_i) + D(w_i).
\end{equation}
Here, $i$ is the index of the cell, $S_i$ is the cell volume, $w_i$ is the vector of conserved
variables, $Q(w_i)$ is the flux of these variables, and $D(w_i)$ denotes a stabilization term. This
evolution equation bears some analogy to the equation of motion used in the Lagrangian SPH method,
c.f. \ref{eqn:SPH_stress_forces}, if the force on a particle is used to define a momentum flux over
a hypothetical interaction area.
The cell-centered Euler CFD scheme allows decoupling between even and odd cells, which leads to
zero-energy modes. The task of the stabilization term is therefore to add coupling by introducing
higher-order terms, namely a second derivative of the conserved variables.:
\begin{equation}
D(w_i) \propto \sum \limits_{n=1}^{NN} \nabla^2 w_i.
\end{equation}
The summation above includes all direct neighbors $NN$ of cell $i$. Specializing to one dimension
only and using central differences, the second derivative is approximated by the undivided
Laplacian $\tilde{\nabla}^2$,
\begin{equation}
	\tilde{\nabla}^2 w_i = w_{l,i} - 2 w_i + w_{r,i},
\end{equation}
where only left and right neighbors with indices $l,i$ and $r,i$ of cell $i$ are considered. Thus,
an explicit and computable expression for the stabilization term is obtained. The interpretation of
this stabilization is that curvature of the solution field is suppressed, i.e., the solution is
forced to be locally linear. The task at hand is now to show that the SPH hourglass control scheme
proposed here can be equally expressed using an undivided Laplacian. To see this, we note from
eqn.~(\ref{eqn:hg_force}), that the hourglass correction force is proportional to the scalar
hourglass error measure, which is a function of the deformation gradient:
\begin{eqnarray}
f_i^{HG} & \propto & \sum  \limits_{j \in \mathcal{S}} \delta_{ij}^i \\
         & \propto & \sum  \limits_{j \in \mathcal{S}} \frac{\left(\bm{F}_{i} \bm{X}_{ij} -
\bm{x}_{ij}\right) \cdot \bm{x}_{ij}} {x_{ij}} \label{eqn:LaplaceProof10}
\end{eqnarray}
For the one-dimensional SPH case, only considering nearest neighbors as well as unit initial
particle spacing and a constant SPH weight function, the deformation gradient is given by Bonet \textit{et al}
\cite{Bonet:2001/a} as $F_i = (x_{r,i} - x_{l,i})/2$, Therefore, eqn.~(\ref{eqn:LaplaceProof10}) is rewritten as
\begin{eqnarray}
f_i^{HG} & \propto & \sum  \limits_{n=1}^{NN} F_i X_{ij} - x_{ij} \\
         & \propto & \sum  \limits_{n=1}^{NN} \frac{x_{r,i} - x_{l,i}}{2} - (x_{r,i} - x_i) \label{eqn:LaplaceProof20} \\
         & \propto & \sum  \limits_{n=1}^{NN} -\left(x_{l,i} - 2 x_i + x_{r,i} \right) \\
         & \propto & \sum  \limits_{n=1}^{NN} -\tilde{\nabla}^2 x_i.
\end{eqnarray}
Note that in eqn.~(\ref{eqn:LaplaceProof20}), the arbitrary substitution $x_{ij} = x_{r,i} - x_i$ has
been made. This analysis shows that the here presented hourglass correction scheme can be expressed
using the undivided Laplacian of the coordinate field. Therefore, the same interpretation of the
stabilization as in the CFD case holds: curvature in is penalized, such that the deformation field
is forced to be locally linear. This interpretation is in agreement with motivation of the SPH
hourglass control scheme, namely that the deformation gradient, which is a linear operator, should
be able to accurately describe the deformation field.


\section{Validation examples}
In the following, a number of test cases is presented which demonstrate the correctness and the
advantages of the hourglass correction scheme presented here. All test cases are computed using 2d
plane-strain conditions and Desbrun's Spiky Kernel \cite{Desbrun:1996/a} is used. It is emphasized
that no artificial viscosity or other dissipative mechanisms are employed to stabilize the
simulations reported here. 

\subsection{Patch test and stability}
\begin{figure}[!ht]
\begin{center}
\includegraphics[width=0.5\textwidth]{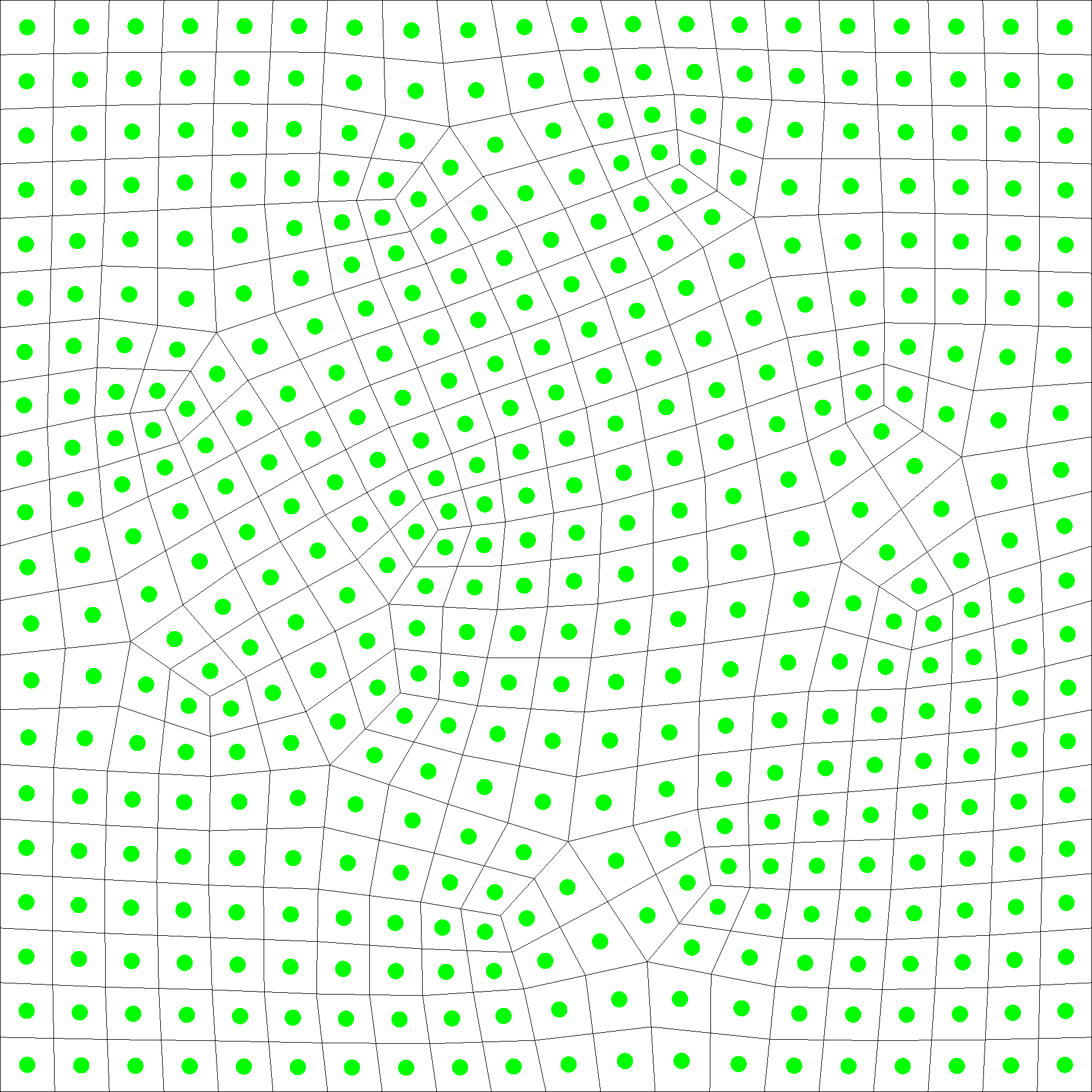}
\end{center}
\caption{Initial particle arrangement for the patch test stability example.
A random initial particle configuration is obtained by discretizing a square patch into 444 quadliteral area elements using a stochastic algorithm based on triangular Voronoi tesselation and recombination of the triangles into
quadliterals. The particle coordinates are taken to be the quadliteral element centers and particle
volumes are taken as the quadliteral area times a unit thickness.}
\label{fig:quad_mesh}
\end{figure}

\begin{figure}[!ht]
\begin{center}
\includegraphics[width=0.99\textwidth]{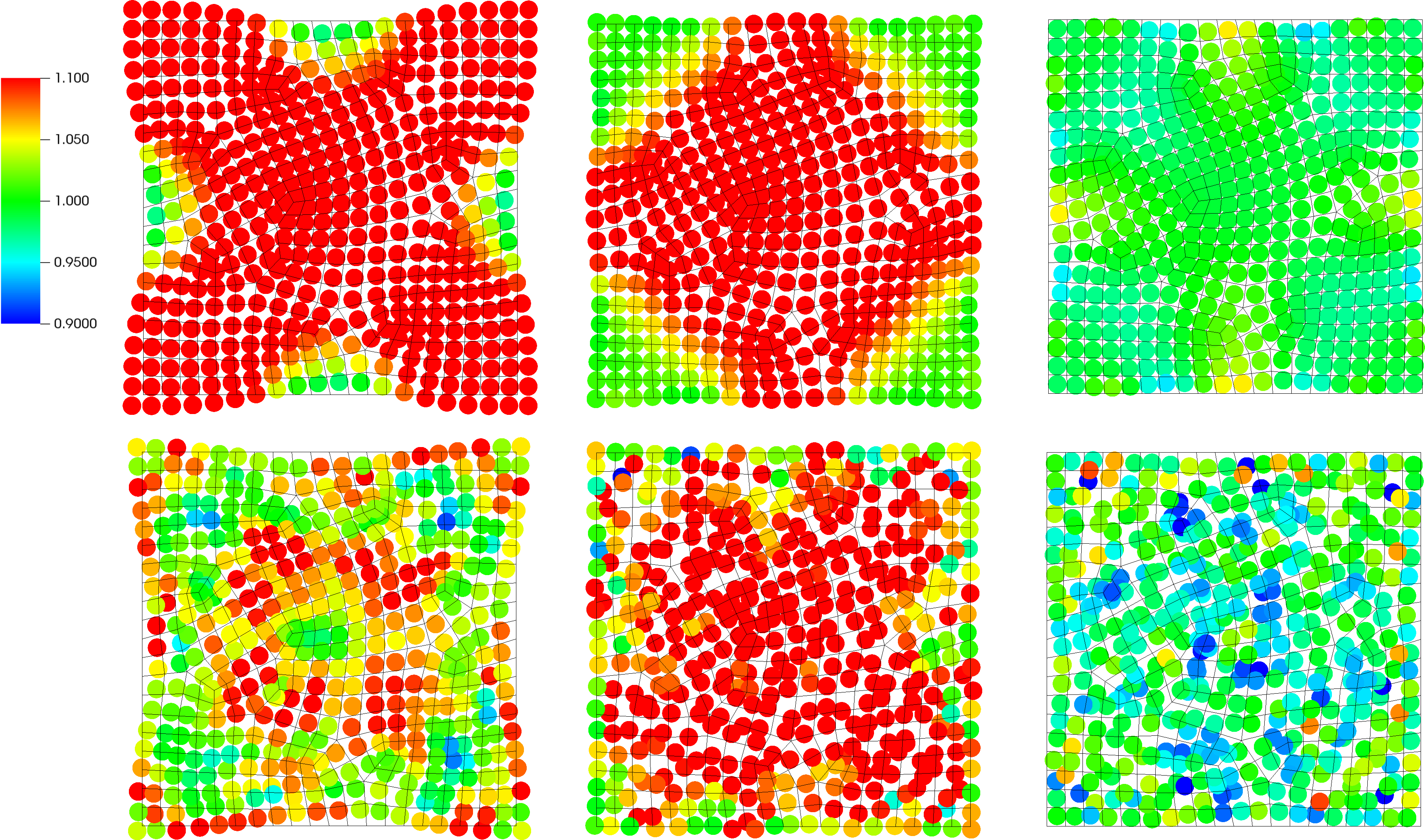}
\end{center}
\caption{Simulation snapshots comparing unstabilized and stabilized trajectories. A 2d patch of an
elastic material is perturbed by isotropic stretch of 10 \% and integrated in time, resulting in
a periodic contraction and expansion mode. The top row of images shows the simulation stabilized
with an hourglass control coefficient $\alpha=50$ during the first, third, and tenth contraction
period. The bottom row shows corresponding snapshots from an unstabilized simulation. The mesh
in the background serves as a guide to the eye.}
\label{fig:quad_patch_comparison}
\end{figure}

The patch test can be regarded as the most basic test for a solid mechanics simulation code. A strip
of some material that is discretized using several elements (or SPH particles) is subjected to a
uniform strain field and the stress is computed. Regardless of the discretization, the stress should
be uniform everywhere. SPH with first-order corrected kernel gradients is well-known to pass this
test. Here, however, we are interested in the subsequent time-integration following such an initial
perturbation. To this end, a
random initial particle configuration is obtained by discretizing a square patch of edge length 1 m and
uniform area mass density $\rho=1 \mathrm{kg/m^2}$ into 444 quadliteral area elements using a stochastic
algorithm based on triangular Voronoi tesselation and recombination of the triangles into
quadliterals. Particle volumes and masses are obtained from the quadliteral area and the prescribed
mass density. The SPH smoothing kernel radius is adjusted individually for each particle such that
approximately 12 neighbors are within the
kernel range. Such a non-uniform particle configuration poses extreme stability challenges to normal
SPH. Figure \ref{fig:quad_mesh} shows the initial particle configuration. The material model is
chosen as linear elastic with $E=1 \:\mathrm{Pa}$ and $\nu=0.3$. The patch is uniformly stretched
by 10\% along both axes. After this initial perturbation, the system is integrated in time
using a standard leap-frog algorithm with a CFL-stable timestep, resulting in a periodic contraction
and expansion mode. Figure \ref{fig:quad_patch_comparison} compares SPH with and without hourglass
stabilization. In the unstabilized system, particle motion is not coherent and particle disorder is
observed already after the first contraction. After eight contractions, particle disorder is very
pronounced and the motion of the patch is entirely dominated by numerical artifacts.
A stabilized simulation was conducted with an hourglass control coefficient $\alpha=50$. This
particular value was found to mark the threshold where a
clear improvement in stability can be observed.
This stabilized simulation exhibits a completely coherent particle motion, and can be continued for
hundreds of oscillations. It is noted that the stabilization increases the apparent stiffness of the
system -- in this case by roughly 3\% as computed from the oscillation frequency. This effect is
similar to hourglass control mechanisms in Finite Element methods, where care has to be taken as to
much hour-glassing control is applied in order to preserve the desired material behavior. This
example is concluded by noting that the here presented hourglass stabilization mechanism leads to a dramatic
increase of the simulation stability even with difficult particle configurations. A video showing
the stabilized and unstabilized simulations is provided as online supplemental data for this
article.

\subsection{Bending of a beam}
\begin{figure}[!ht]
\begin{center}
\includegraphics[width=0.75\textwidth]{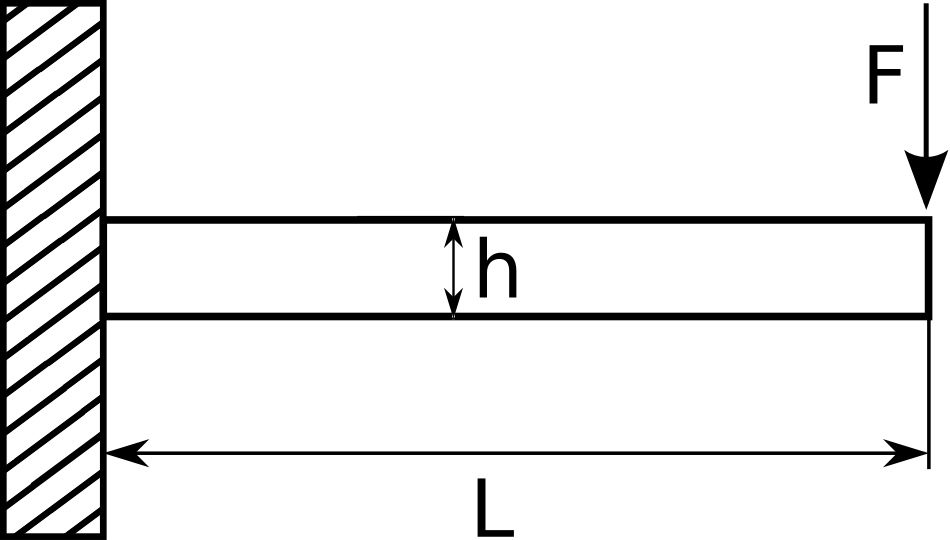}
\end{center}
\caption{Sketch of the simply supported beam with dimensions $L=8\,\mathrm{mm}$ and $h=1\, \mathrm{mm}$.}
\label{fig:beam_sketch}
\end{figure}

\begin{figure}[!ht]
\begin{center}
\includegraphics[width=0.99\textwidth]{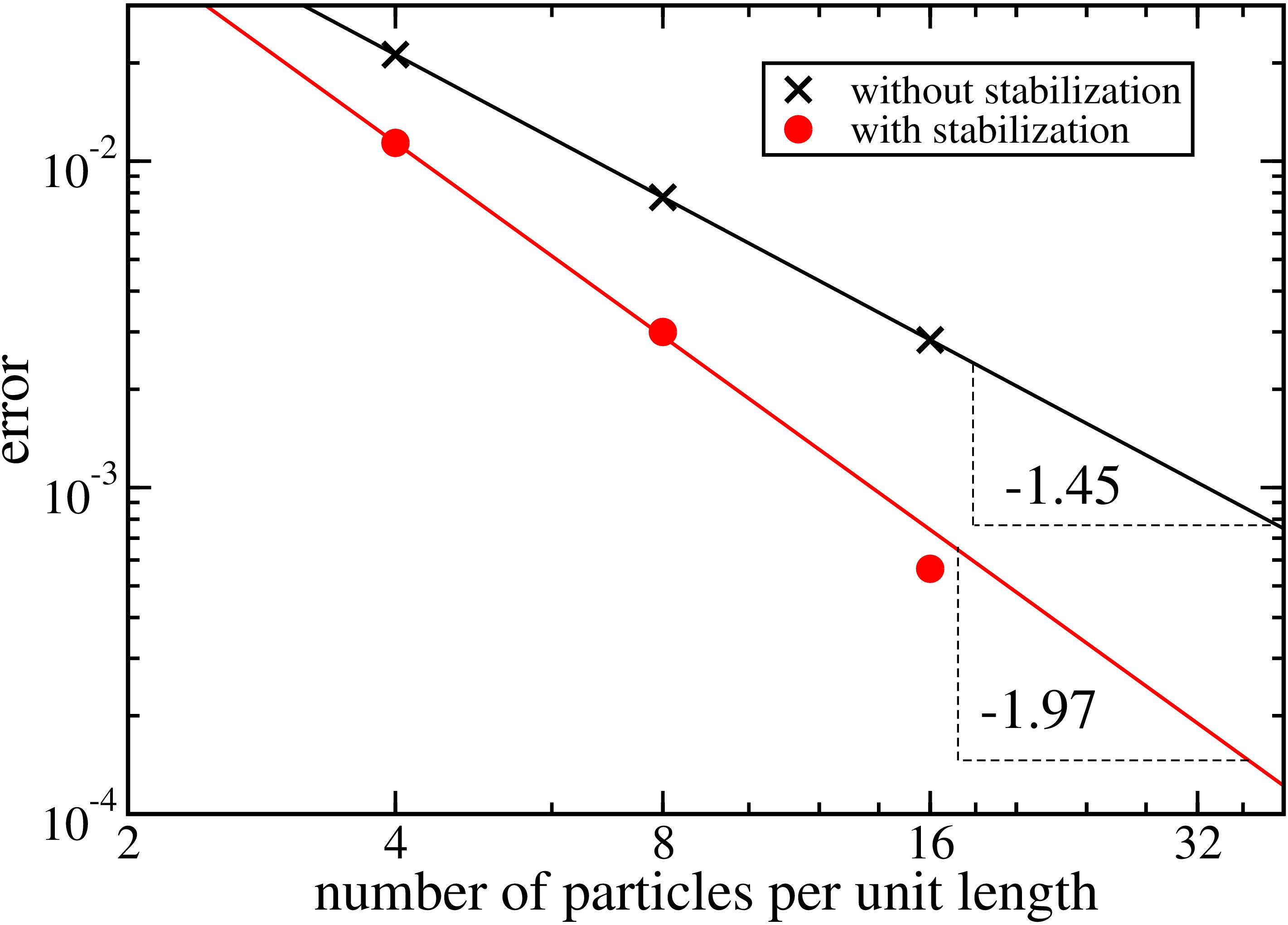}
\end{center}
\caption{Relative errors of the SPH solution to the bending problem for the simply supported beam.
The unstabilized solution employs hourglass control with $\alpha=50$, while the unstabilized solution uses
$\alpha=0$.}
\label{fig:beam_convergence}
\end{figure}

This example tests the accuracy of the stabilized SPH scheme against the analytic Timoshenko
solution to the bending of a simply supported beam with dimensions $8\,\mathrm{mm} \times
1\,\mathrm{mm}$, see Fig.~\ref{fig:beam_sketch}.The material model of the beam is taken as linear elastic with Youngs modulus $E=100\,\mathrm{GPa}$ and $\nu=0.3$. As this is a 2d plane strain problem, the effective 2d Youngs
modulus is
\begin{equation}
	E_{2d} = \frac{E}{1 - \nu^2} \simeq 109.89 \,\mathrm{GPa}
\end{equation}
Timoshenko's solution \cite{Timoshenko:1970/a} for this problem results in a compliance (the ratio
of applied displacement to the resulting force) $C_{theory}=f/u\simeq0.053657\, \mathrm{kN}/\mathrm{mm}$.
For the numerical SPH solution, the beam is discretized using a 2d square lattice with a total
number of particles $N = n_x \times n_y = (3+8n) \times n$, with $n \in \left[ 4, 8, 16 \right]$.
The smoothing kernel radius is chosen as 2.41 times the lattice spacing, such that the number of
neighbors is constant for all discretizations.  The beam is clamped at the left end by enforcing zero
displacements for the three leftmost columns of particles. A downwards displacement boundary
condition of maximum amplitude $u_y=0.1\,\mathrm{mm}$ is applied  at the top right particle with a
velocity of 0.1 mm/s which corresponds to quasi-static loading.  The reaction force $f$ is
measured at this particle, resulting in the numerical compliance $C_{num}$. To quantify the accuracy
of the numerical solution, the $L^2$-norm of the relative error between the numerical solution and
the analytic solution is defined as
\begin{equation}
	\mathrm{error} = \sqrt{ \left( \frac{C_{num}-C_{theory}}{C_{theory}} \right)^2 }.
\end{equation}
Fig.~\ref{fig:beam_convergence} reports error values for an unstabilized simulation and a stabilized
simulation for different discretization length scales. For the unstabilized simulation, the
convergence rate is approximately -1.45. It is not clear how to understand this odd rate of
convergence, as the Lagrangian SPH method employed here uses a kernel gradient approximation which
is correct to first order. Therefore, all linear terms in the solution should be computed exactly
such that only terms of second order terms contribute to the error, and one might speculate that
integration errors due to nonlinear modes increase the errors. In contrast to this behavior, we
observe that the stabilized simulation yields almost exactly quadratic convergence as well as much
lower relative error values. This result is remarkable, as -- to the best of this author's knowledge
-- no other nodal integration scheme is able to yield quadratic convergence, and only
computationally expensive cell integration methods have been reported to perform in this manner
\cite{Rabczuk:2004a}. A parameter study of the hourglass control coefficient reveals that the factor
$\alpha=50$, which was observed to mark the threshold beyond which the patch test simulation from
the previous section was clearly stabilized, is also a good choice for the beam bending problem. As
can inferred from Table \ref{table:parameter_study}, which lists the convergence rates and the
relative errors of the simulation with the finest discretization for different values of $\alpha$
are listed, the convergence rate and error magnitude is nearly optimal for $\alpha=50$. The author
is aware that other optimal values might be determined for different simulation problems, however,
for the limited scope of this work, $\alpha = 50$ is used henceforth for all stabilized simulations.

\begin{table}
\begin{center}
\caption{Parameter study for the hour-glassing control coefficient $\alpha$ and the bending problem of
a simply supported beam. The rate of convergence
refers to the power law exponent which describes how the numerical solution approaches the
analytic solution upon discretization refinement. The relative error quotes the magnitude of
deviation between the finest discretization and the analytic solution.}
\label{table:parameter_study}
\begin{tabular}{ l | c | r }
  $\alpha$ & convergence rate & error at finest discretization\\
  \hline                       
  0 & -1.45 & $24.5 \times 10^{-4}$ \\
  25 & -1.65 & $18.7 \times 10^{-4}$ \\
  50 & -1.97 & $6.6 \times 10^{-4}$ \\
  75 & -1.78 & $21.9 \times 10^{-4}$ \\
  100 & -1.74 & $31.7 \times 10^{-4}$
\end{tabular}
\end{center}
\end{table}


\begin{figure}[!ht]
\begin{center}
\includegraphics[width=0.99\textwidth]{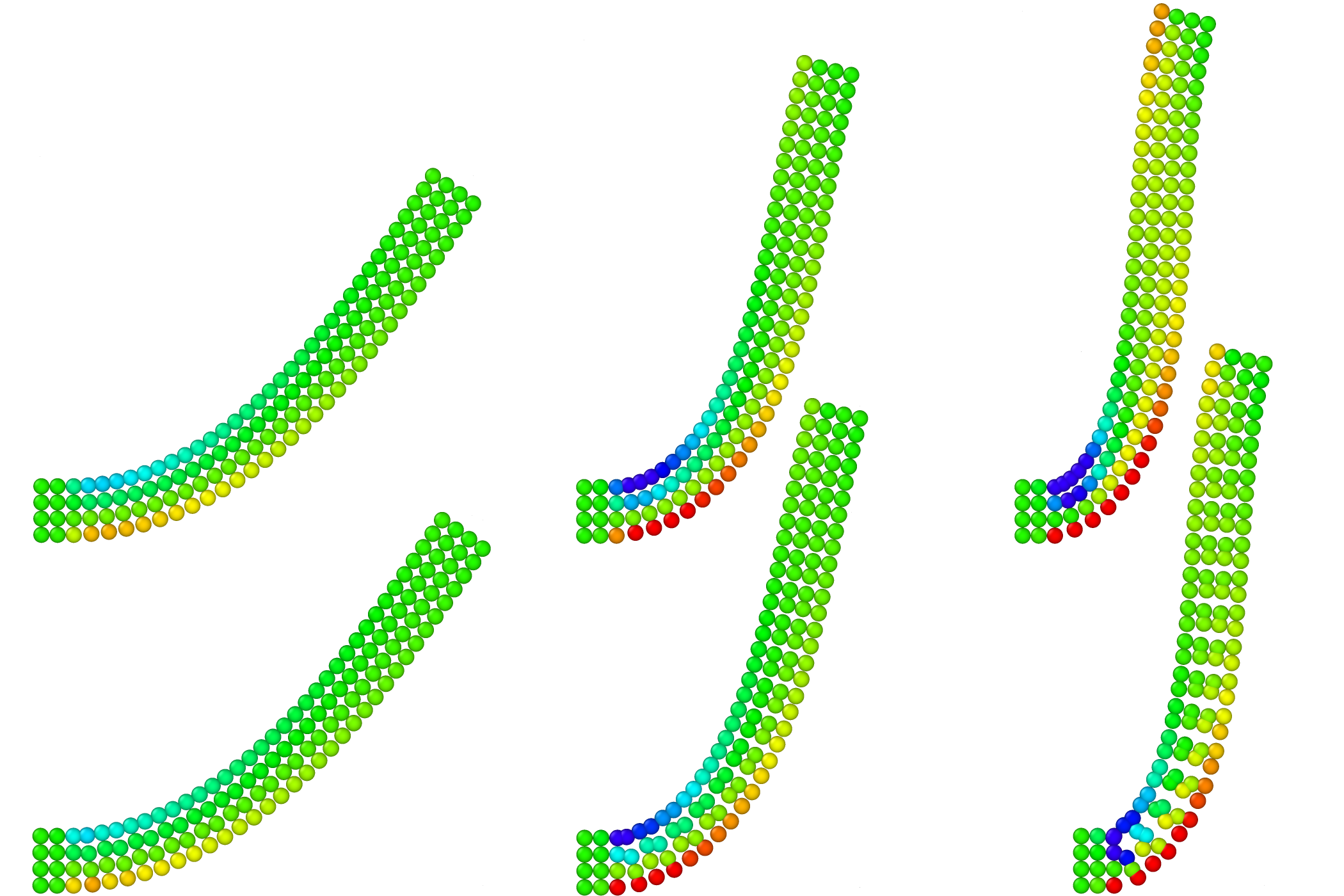}
\end{center}
\caption{Comparison of stabilized (top) and unstabilized results for a large deflection simulation
of the simply supported beam. The color coding represents the $xx$-component of the Second
Piola-Kirchhoff stress tensor, with red signifying tension and blue compression.}
\label{fig:beam_comparison}
\end{figure}

The above test case has only considered a maximum beam deflection which is small compared to
the beam dimensions for the purpose of comparing against an analytic solution. It is instructive
to extend this example to very large deflections. To this end, a large boundary displacement
$u_y=8\,\mathrm{mm}$ is applied, and stabilized and unstabilized simulation results are
qualitatively compared. Snapshots of the simulations corresponding to the same boundary
displacements, are shown in
Fig.~\ref{fig:beam_comparison}. In the first snapshot at $u_y=4.8\,\mathrm{mm}$, no obvious
difference between both simulations can be observed. For the second snapshot at
$u_y=6.6\,\mathrm{mm}$, the unstabilized simulation shows a pattern where pairs of particles in the
interior of the beam move together and do not follow the rotation of the beam's neutral line.
Clearly, the unstabilized system tries to minimize its elastic energy by assuming a configuration
which is not in agreement with a linear displacement field. This leads to an obviously wrong
particle configuration for the last snapshot at $u_y=8.0\,\mathrm{mm}$. In contrast, the stabilized
system with $\alpha=50$ shows particle displacements which appear very reasonable and exhibit no
instability.
We conclude this test case by noting that Total-Lagrangian SPH can become unstable for certain
loading conditions even with perfectly uniform reference particle configurations. The here proposed
hourglass-control algorithm removes these instabilities and increases accuracy as well as
convergence rate.


\subsection{Punch Test} \label{sec:punch_test}
The punch test is a classic example for a difficult large deformation test. It was first applied to
a meshless method in order to demonstrate the stability of the Reproducing Kernel Particle
Method, which combines kernel approximation techniques with Gaussian integration on a background
mesh \cite{Chen:1996/a}. In the punch test, a block of a rubber-like material is compressed by a
rigid rectangular tool with sharp edges. As the compression proceeds, the material moves outwards
to open sides. The punch tool digs into the material and causes large stress concentrations at its
sharp corners. This problem is difficult to simulate with FEM and impossible with unstabilized
Lagrangian SPH \cite{Vidal:2007/a}. The rubber material is described by a compressible Neo-Hookean
model \cite{Bonet:2008/a}, 
\begin{equation}
	\bm{S} = \mu \left( \bm{I} - \bm{C}^{-1} \right) + \lambda (\ln J) \bm{C}^{-1},
\end{equation}
where $\lambda=E \nu / [(1+\nu)(1--2\nu)]$ and $\mu=E/[2(1+\nu)]$ are the Lam\'e constants, $J$ is
the determinant for the deformation gradient $\bm{F}$, and $\bm{C}=\bm{F}^T \bm{F}$. The rubber is
given by a rectangular block of dimensions $6\,\mathrm{mm} \times 3\,\mathrm{mm}$, which is
discretized using a square lattice with lattice constant $l_0=1/15 \mathrm{mm}$. The material
parameters Young's modulus and Poisson ratio are taken as $1\,\mathrm{GPa}$ and 0.45, respectively.
The compression tools are modelled as thin blocks of a rigid material with dimensions
$6\,\mathrm{mm} \times 0.15\,\mathrm{mm}$, discretized as a square lattice of particles with lattice
constant $l_1=l_0/4$. The tool particles interact with the rubber particles via a repulsive Hertzian potential
\cite{Landau:1986/a} which utilizes the same Young's modulus as the rubber and is effectively so
stiff that no mutual penetration can occur.
\begin{figure}[!ht]
\begin{center}
\includegraphics[width=0.99\textwidth]{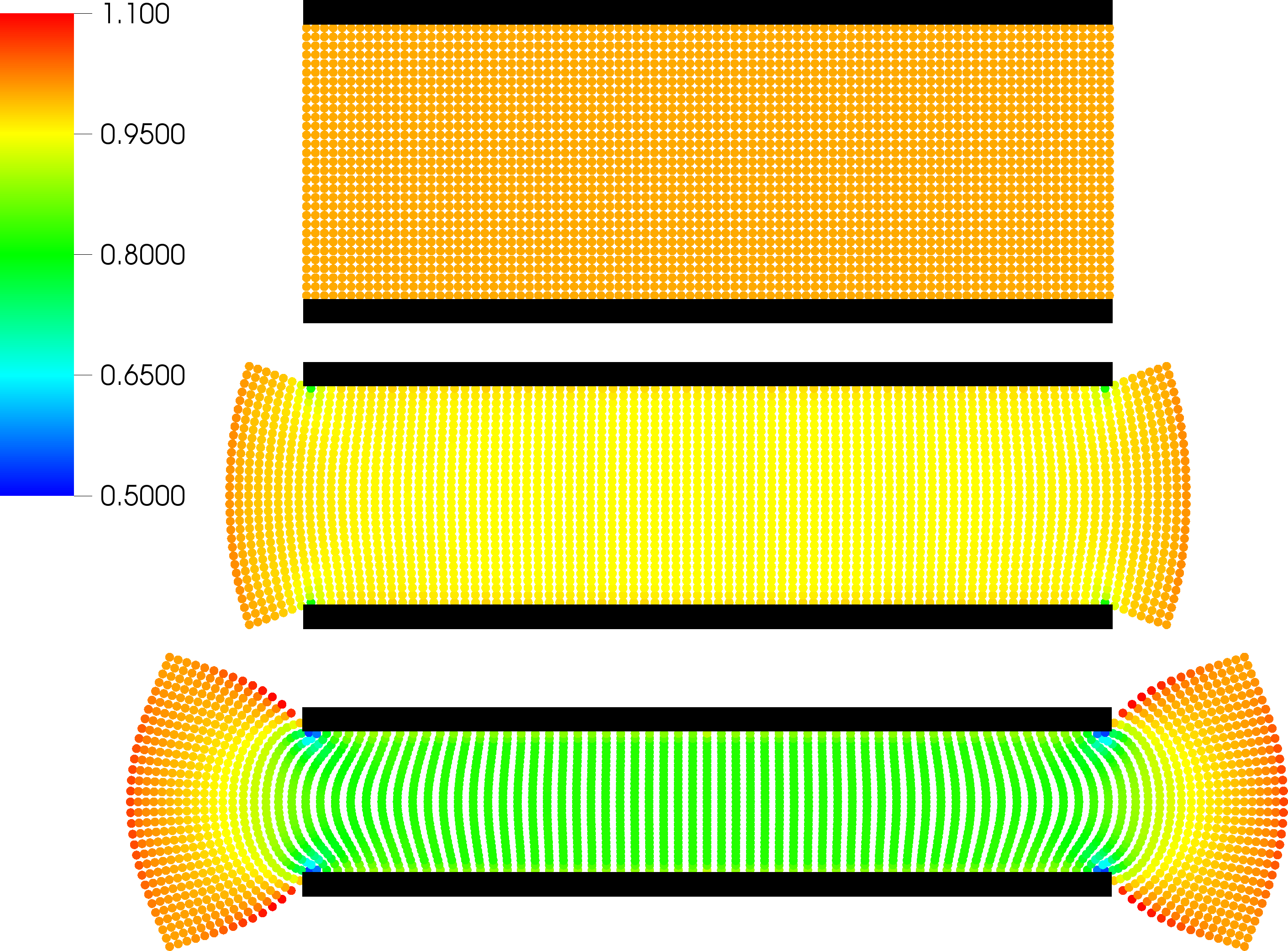}
\end{center}
\caption{Snapshots of the punch test simulation corresponding to vertical compression states of 0\%,
10\% and 50\%. Rubber particles are colored according to the determinant of the deformation
gradient. Note the strong gradient of volumetric strain at the corners of the compression tool
plates, which are shown in black.}
\label{fig:punch_test}
\end{figure}

The simulation is performed with a compression velocity of 1 mm/s, which can be considered
quasi-static, and a stabilization parameter $\alpha=50$. Simulation snapshots are shown in
Fig.~\ref{fig:punch_test}. The simulation,proceeds without any sign of instabilities until a vertical
compression ratio of 50\% is reached. Higher velocities up to 100 mm/s and further compression up to
90\% were also simulated, however, strongly dynamic compression rates necessitate the use of
artificial viscosity to keep the simulation stable. This test case is concluded by noting that an
otherwise difficult simulation problem can be easily handled using Total-Lagrangian SPH stabilized
with the hourglass control mechanism presented here.


\subsection{Rubber Pull Test}
\begin{figure}[!ht]
\begin{center}
\includegraphics[width=0.99\textwidth]{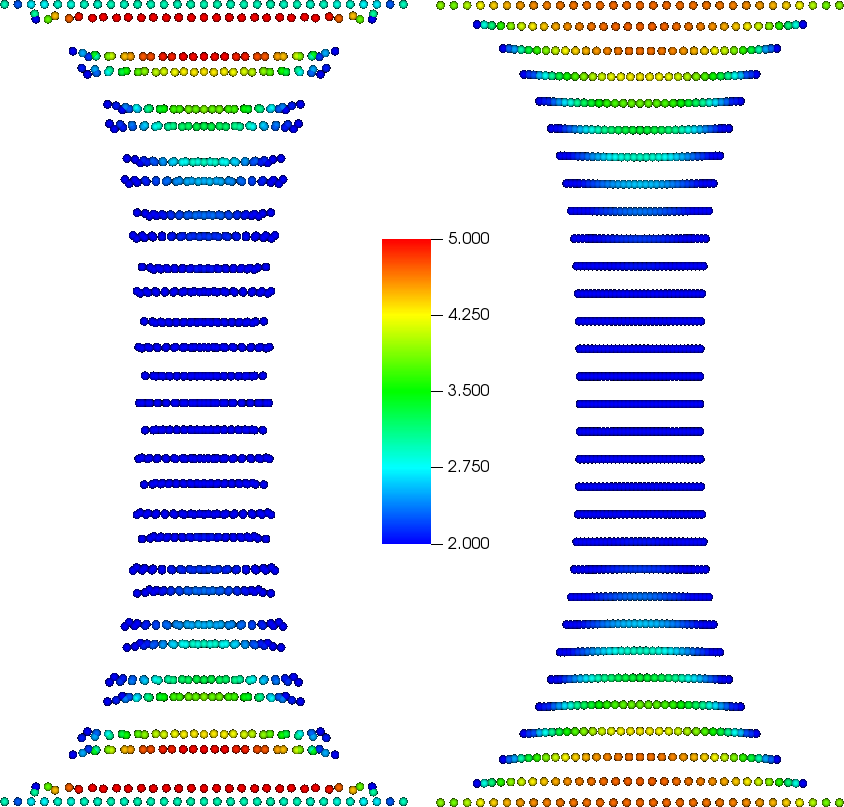}
\end{center}
\caption{Comparison of unstabilized (left) and stabilized (right) simulation snapshots for the
rubber strip pull test case. An initially quadratic rubber strip is elongated in the vertical
direction until a strain of 500\% is reached. The snapshots are scaled here in the vertical
direction to better fit them on the page. Color coding of the particles represents the determinant
of the deformation gradient.}
\label{fig:rubber_pull}
\end{figure}

In this example, a strip of rubber is subjected to a 500\% increase in length. This test case is
also taken from \cite{Chen:1996/a}, where it serves as a very large deformation test to demonstrate
the stability of the RKPM method. The rubber strip is initially quadratic with dimensions of
$2\,\mathrm{mm} \times 2\,\mathrm{mm}$ and discretized using a square lattice with lattice constant
$l_0=1/15\,\mathrm{mm}$. The smoothing kernel radius is taken as $2 l_0$. The top and bottom row of
particles are subjected to a velocity boundary condition $\bm{v}=(0,\pm 0.1 \mathrm{mm/s})$ which
causes quasi-static elongation of the rubber strip, while no lateral displacement is possible for
these rows of particles. This setup corresponds to the clamping condition typically encountered in
material test stands. The same compressible Neo-Hookean material model is used as in the previous
test case. The simulation is carried out until the length of the rubber strip has increased to
$12\,\mathrm{mm}$.  Fig.~\ref{fig:rubber_pull} compares the final state of an unstabilized
simulation with that of a simulation stabilized with $\alpha=50$. The unstabilized simulation
exhibits particle disorder, in particular near the boundaries of the domain. A pairing of rows of
particles, alternating with larger gaps, similar to the pairing effect seen in the large deflection
simulation of the simply supported beam, can be observed. No such instabilities are present in the
stabilized simulation, which features a very smooth deformation field.

\subsection{Taylor impact}
All of the preceding test cases have investigated elastic material behavior only, without any kind
of material instability. The deformation field for an elastic material is typically very smooth, in
contrast to a plastic deformation, which is governed by the localization of deviatoric stress.  This
localization is usually well defined with a sharp boundary, e.g., in the case of adiabatic shear
bands. To successfully cope with plastic deformations, the numerical solution scheme must therefore
allow some desired nonlinearities to develop, while other nonlinearities, e.g. zero-energy modes,
must be suppressed. This requirements poses a severe challenge to the hourglass-control algorithm
presented here. Let us briefly remind ourselves what the effect of this algorithm is: the
deformation associated with a line element in the current configuration is compared against the result of using
the deformation gradient (which is a linear operator) on a line element in the reference configuration. Deviations between both
deformations are minimized by a penalty force. This means that the deformation field is forced to be
locally (on the scale of the smoothing kernel radius) linear. It is therefore expected that
localization phenomena such as plastic deformation are suppressed by the hourglass control mechanism.
This behavior is again similar to what is observed for hourglass control schemes in FEM: there, the
amplitude of the control force has to be scaled down by orders of magnitude in order to permit
plastic deformation, or indeed fluid flow, to happen.

With this in mind, the SPH hourglass control mechanism is reformulated as follows. All SPH particles
with a zero value of plastic deformation utilize the expression already given by
eqn.~(\ref{eqn:hg_force}).  For any SPH particle with a non-zero value of plastic deformation, the
Young's modulus in eqn.~(\ref{eqn:hg_force}) is replaced by the yield strength, which is typically
much smaller than the Young's modulus.

As a plastic deformation test case, the impact of an aluminum bar against a rigid plate is considered.
The material parameters are given by Young's modulus $E=\mathrm{78.2\ GPa}$, yield strength
$\sigma_Y=1.0\mathrm{\ GPa}$, Poisson ratio $\nu=0.3$, and mass density $\rho=\mathrm{2700\ kg\,
m^{-3}}$. The aluminum bar travels at an initial velocity of $v_0=\mathrm{373\ m\,s^{-1}}$ and has
dimensions of $3.91\,\mathrm{mm} \times 23.46\,\mathrm{mm}$. It is discretized using a square lattice
of particles with a lattice spacing of $l_0=0.35\,\mathrm{mm}$. The smoothing kernel radius is taken
as $3 l_0$. The material model used here is linear elastic -- perfectly plastic and based on the
multiplicative decomposition of the total deformation gradient in elastic and plastic parts, which is
described in detail in Appendix A of \cite{Bonet:2008/a}.

\begin{figure}[!ht]
\begin{center}
\includegraphics[width=0.99\textwidth]{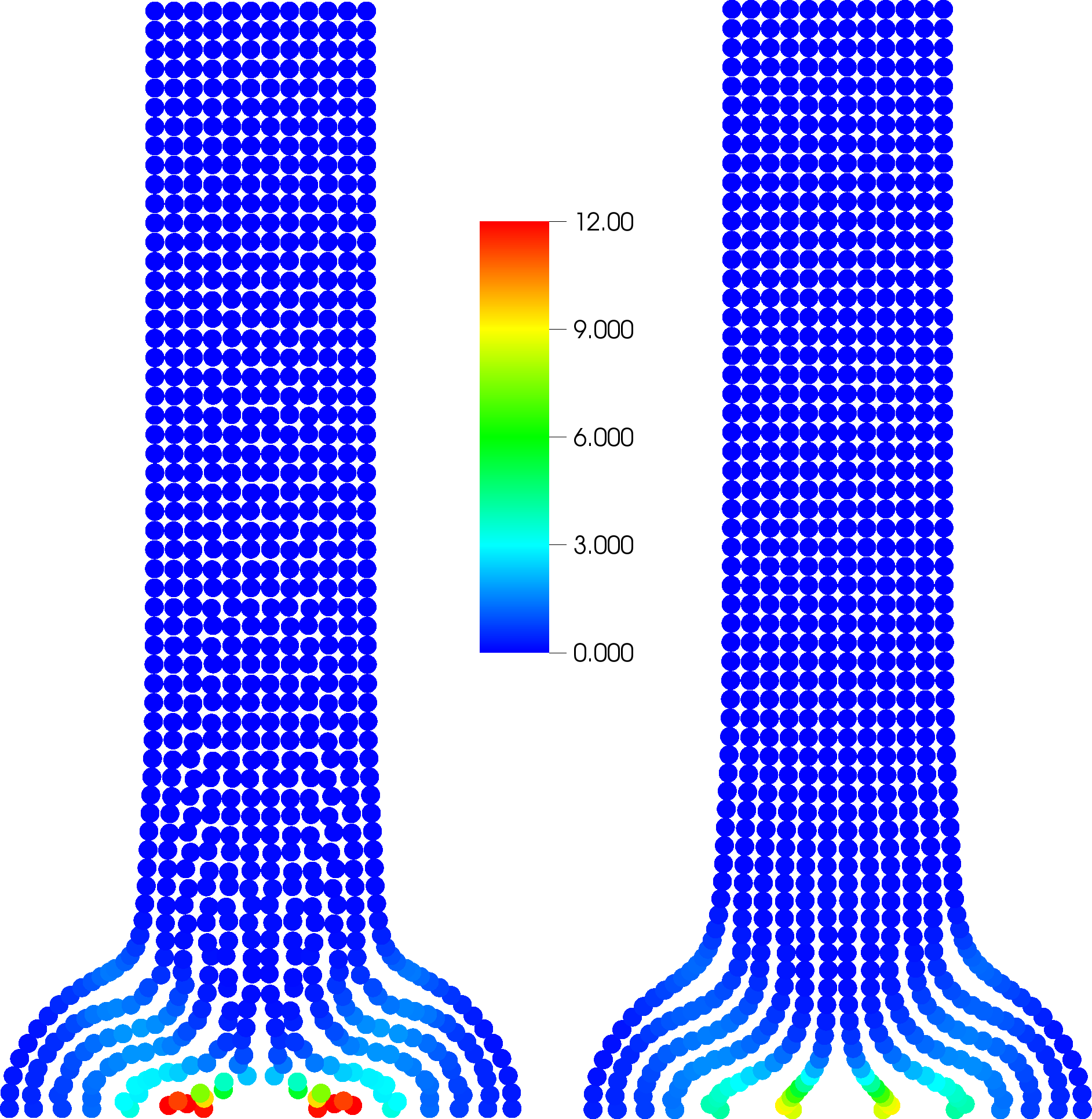}
\end{center}
\caption{2d Taylor impact test. The material model is linear-elastic / ideal plastic. Color coding
is the norm of the plastic strain tensor. The left image shows the unstabilized SPH solution with
clearly visible particle disorder. In contrast, SPH stabilized with the here proposed algorithm
results in a smooth solution.}
\label{fig:taylor_impact}
\end{figure}

The simulation is carried out until the development of plastic deformation is complete.
Fig.~\ref{fig:taylor_impact} compares the unstabilized solution against the stabilized solution.
The general shape of the deformation appears reasonable, and is similar to published results for the
2d axissymetric case \cite{Chen:1996/a}, even though the 2d plane-strain case is considered here.
The unstabilized simulation exhibits particle disorder, which is particularly pronounced in the
regions of strong plastic deformation, but also extends into regions with only elastic deformation.
We note that the addition of a reasonably small amount of artificial viscosity would prevent the
occurrence of particle disorder. For the purpose of demonstrating the onset of instability, however,
we chose to report results with no artificial viscosity.  It is likely, that all SPH results
published for this and similar problems use dissipative mechanisms to keep the solution smooth.  In
the case of the stabilized simulation, a smooth deformation field is obtained  without the need for
artificial viscosity. The overall shape of the deformed aluminum bar is similar to the unstabilized
solution, indicating that plastic flow was not excessively suppressed by the stabilization
algorithm. It is noted, however, that if no reduction of the hourglass control amplitude as
discussed above is performed, plastic deformation does not occur and the aluminum bar rebounds
elastically from the rigid surface.

This example is concluded by noting that the SPH hourglass control algorithm presented here
may be used to simulate material instabilities such as plastic deformation, but care has to be taken
as to how much hour-glassing control is applied.

\subsection{Tensile test of a 3d specimen with failure}
\begin{figure}[!ht]
\begin{center}
\includegraphics[width=0.99\textwidth]{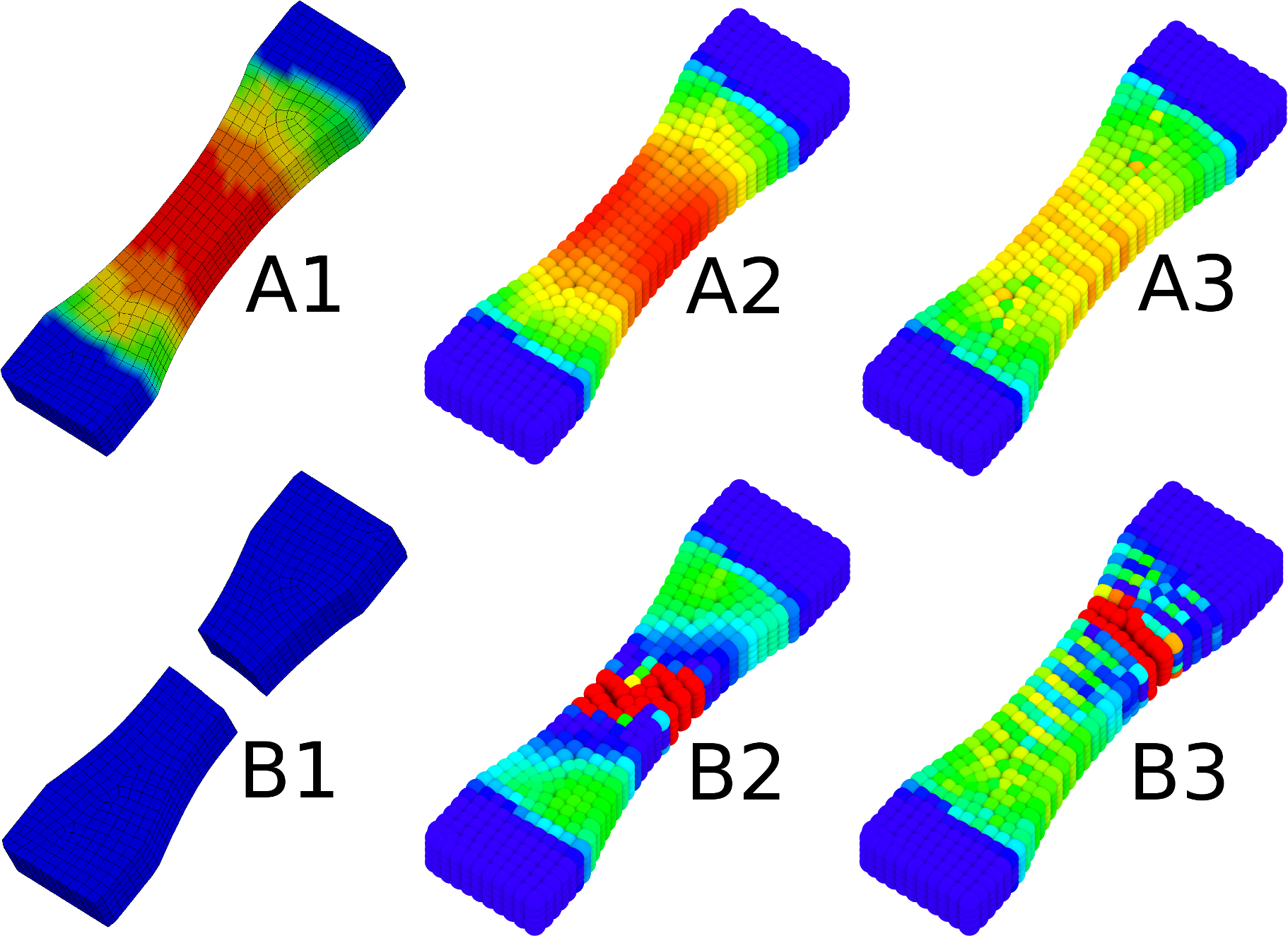}
\end{center}
\caption{Comparison of stabilized and unstabilized SPH with FEM. The test specimen with a gauge
cross section of $15\times10\, \mathrm{mm}$ is discretized for FEM using 1800 solid elements. The SPH mesh is
derived by placing a particle at the barycenter of each element, and choosing the kernel radius
such that 30 neighbors are obtained. Images A1 and B1 show the FEM solution just before and after
failure. A2/B2 show the same snapshots for the stabilized SPH simulation, while A3/B3 show the
unstabilized SPH simulation. The color coding respresents maximum principal strain, increasing from
blue to red}
\label{fig:tensile3d}
\end{figure}

The purpose of this test example is to compare the failure behavior of SPH with the here presented
stabilization algorithm against unstabilized SPH and a FEM solution (reduced integration elements
with one Gauss point). A 3d tensile test specimen is
elongated at $10\, \mathrm{mm}/\mathrm{ms}$ until failure occurs. The material
model used is linear-elastic with  $E=\mathrm{10\ GPa}$, $\nu=0.3$, and $\rho=\mathrm{2700\ kg\,
m^{-3}}$. For SPH, failure is realized with a damage rate model similar to \cite{Benz:1995/a},
which accumulates damage if any current principal strain value exceeds a pre-defined value $\epsilon_{max}=0.08$:
\begin{equation}
	\frac{\mathrm{d}D^{1/3}}{\mathrm{d}t} = 0.4\, \frac{c_0}{h}
\end{equation}
In the equation above, $0.4\, c_0$ is the damage propagation speed which ensures that a crack does
not propagate faster than the speed of sound $c_0$ over a characteristic distance $h$, here taken as the
SPH kernel radius. The scalar damage variable is used to scale the components $\sigma_i$ of the principal (diagonalized) stress
tensor only if the component is positive, i.e., damage is only allowed for tension and not for
compression:
\begin{equation}
	\tilde{\sigma}_i = (1 - D)\, \sigma_i
\end{equation}
The scaled stress tensor $\tilde{\sigma}$ is then used to compute interparticle forces.  In the case
of FEM, failure is realized by instantaneously eroding those elements which exceed the maximum
principal strain.  Fig.~\ref{fig:tensile3d} compares snapshots of stabilized and unstabilized SPH
solutions with the FEM solution. The comparison indicates that stress distribution of the stabilized
SPH solution is as smooth as the FEM solution and concentrates at the center of the test specimen,
where the minimum cross section is located. In contrast, the stress distribution of the unstabilized
SPH solution is affected by the non-uniform particle spacing and exhibits alternating stripes of low
and high stress values. As a result, failure is incorrectly predicted at an off-center position. The
stabilized SPH solution, however, correctly predicts failure at the center, in agreemeent with FEM.

\begin{figure}[!ht]
\begin{center}
\includegraphics[width=0.99\textwidth]{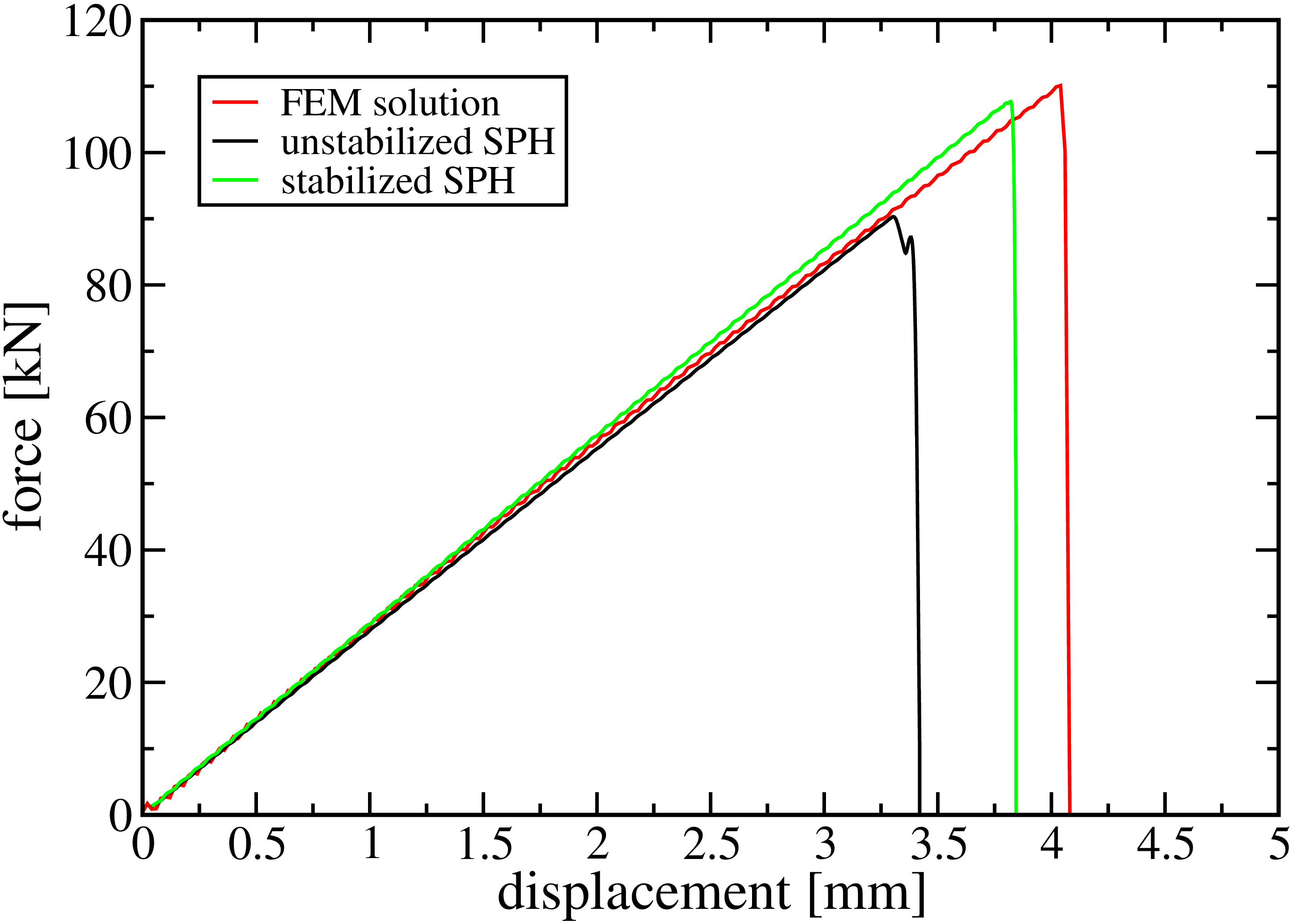}
\end{center}
\caption{Force/displacement data for the 3d tensile test. 
Comparison is made between stabilized and unstabilized SPH, and a FEM reference solution.}
\label{fig:tensile3d_graph}
\end{figure}

It is also of interest to quantitatively compare the reaction force at the clamped ends of the
tensile specimen for the three different simulation approaches. Fig.~\ref{fig:tensile3d_graph} shows
that stiffness obtained with SPH agrees very well with the FEM solution. However, the unstabilized
SPH solution predicts failure at a displacement value which is 15\% smaller than the FEM solution.
In case of the stabilized SPH solution, this value decrease to 4\%. The comparison therefore
suggests that Total Lagrangian SPH stabilized with the here presented algorithm yields a similar accuracy as FEM with
reduced integration elements, if the discretization is maintained at a similar level, i.e., elements
are replaced with particles on a one-to-one basis.

\section{Discussion}
This paper presents a stabilization scheme for Smooth-Particle Hydrodynamics (SPH). The
Total-Lagrangian formulation of SPH is considered, which does not suffer from tensile instability
\cite{Swegle:1995/a}, but is still affected by another instability caused solely by the collocation
character: rank-deficiency \cite{Dyka:1997/a}. The mathematical meaning of rank-deficiency is that
the solution to the underlying equilibrium equations which are discretized with SPH is non-unique. A
physical interpretation of this non-uniqueness is obtained by considering particle positions and the
deformation field. Within the kernel radius of a particle, there can be a manifold of different
arrangements of particles which all result in the same SPH approximation of the deformation gradient.
The deformation gradient results in a strain measure, which, via the material model, results in a
stress state with an associated elastic energy.  Oscillatory modes between the non-unique
arrangements of particles cannot be suppressed, because the states which are linked by the
oscillations are not distinct in terms of elastic energy. Thus, these oscillations modes can be
populated with arbitrarily high amounts of kinetic energy until the solution is entirely dominated
by this artifact, such that the simulation can be essentially considered as failed.

Various methods have been proposed to suppress these modes. Of particular importance are the addition
of stress points \cite{Dyka:1997/a, Randles:1996a}, which eliminate the rank-deficiency at the cost
of an additional integration overhead. While this approach is very clean and attractive from a
mathematical perspective, it appears somewhat impractical to use and not many researchers have taken
up this approach, as judging from the available literature. An alternative approach is to suppress
the growth of the instability modes by simply damping them out. This can be conveniently realized
via the use of artificial viscosity , or conservative smoothing \cite{Monaghan:1989/a,
Hicks:1997/a}. However, by doing so, the simulation typically becomes more dissipative or dispersive
in nature as one would like it to be.

This article introduces a completely different and new method to suppress the instability modes. The
method is inspired by stabilization algorithms used in a  non-particle method, namely the
Finite-Element Method (FEM), which exhibits a very similar instability if rank-deficient elements
are employed. In FEM, this instability is termed hour-glassing, and it is counteracted by suppressing
non-linear deformation modes \cite{Flanagan:1981/a}. This work adapts this approach by identifying
suppressing all modes which are not described by the deformation gradient. As the deformation
gradient is a linear operator, a linear deformation field is enforced, and, at the same time, a
unique correspondence between pairs of particle displacements and the deformation gradient is
established.

The performance of the resulting stabilization algorithm is assessed using a number of linear
elastic and elastoplastic test cases. In the case of the bending of a beam, a very desirable result
is obtained: For a certain range of the amplitude of the stabilization force, quadratic error
convergence is observed. This result is remarkable as it indicates that the stabilized numerical
solution treats all linear terms correctly, such that remaining errors are solely of second order.
This is precisely what one would expect from a numerical scheme which is first-order correct, such
as gradient-corrected SPH. However, results reported for unstabilized SPH show convergence behavior
which is less than quadratic \cite{Rabczuk:2004a}, presumably because numerical artifacts affect the
solution. Thus, the stabilization scheme reported here provides a clear improvement.

In the case of large rotation and deformations, much improved stability behavior is observed when
compared to unstabilized SPH. The stabilized solution is always smooth, even in the case of the
difficult punch test. In contrast, unstabilized SPH exhibits instabilities which are typically
characterized by pairwise clumping of particles. Improved stability is also obtained for plastic
deformations, however, the amplitude of the stabilization algorithm must be carefully chosen in this
case, as too much stabilization inhibits plastic flow. This behavior is very similar to hour-glassing
control mechanisms in FEM, where the amount of hour-glassing control typically has to be scaled down
by orders of magnitude for plasticity problems when compared to purely elastic deformations
\cite{CesardeSa:2001/a}. This behavior is not ideal, but the here presented conditional switch,
which reduces the stabilization strength at the onset of plastic deformation, appears to work to a
satisfactory degree. Future work along these lines should address a viscous -- rather than stiffness
-- based formulation of the stabilization algorithm.

It is noteworthy to discuss another stabilization algorithm for Total-Lagrangian SPH. Vidal \textit
{et al.} developed an approach, which minimizes a local measure of the Laplacian of the deformation
field. Their approach results in much enhanced stability, such that updates of the reference
configuration can be performed, yielding a stabilized updated Lagrangian scheme. Their work is
seminal in the sense, that (i) updates of the reference configuration were identified to be a cause
for the occurence of zero-energy modes, and (ii) higher-order derivatives could be used to suppress
such instabilities. However, the algorithm involves the computation of a third-order tensor, which
-- in this author's experience -- dramatically affects the numerical efficacy. Additionally, the
stabilizing effect is not very clear for large deformations, as the authors report that the punch
test (c.f. Sec. \ref{sec:punch_test}) cannot be performed without updates of the reference
configuration. In contrast, the stabilization scheme reported here, which is based on the
deformation gradient bears negligible impact on the speed of the simulation.  Early results of
applying the new stabilization scheme to the updated Lagrangian formalism indicates that it also
works very well, however, these results are outside the scope of the current article and will be
published elsewhere.

What are the implications of this work? The stabilization algorithm reported here dramatically
extends the range of deformations that can be simulated with Total-Lagrangian SPH in a stable and
manner, without resorting to highly dissipative mechanisms. The new method is not affected by
non-uniform particle configurations, as shown by the patch test. As the deformation field remains
smooth, even for large deformations, numerical artifacts are less of an issue compared to standard
SPH, such that higher accuracy is obtained. This statement is supported by the enhanced error
convergence rate observed in the beam bending problem. Because the new stabilization algorithm bears
analogies to hour-glassing control for FEM, the effect of the additionally introduced stiffness can
be understood and judged using the large body of literature published for that method.  The current
method is straightforward to implement, and does not incur any significant computational overhead.
It is therefore expected that this contribution pushes SPH away yet another step away from its
academic status towards a method that can actually be used in practice for the reliable simulation
of large deformation problems.


\begin{thebibliography}{10}
\expandafter\ifx\csname url\endcsname\relax
  \def\url#1{\texttt{#1}}\fi
\expandafter\ifx\csname urlprefix\endcsname\relax\def\urlprefix{URL }\fi
\expandafter\ifx\csname href\endcsname\relax
  \def\href#1#2{#2} \def\path#1{#1}\fi

\bibitem{Lucy:1977/a}
L.~B. Lucy, A numerical approach to the testing of the fission hypothesis, The
  Astronomical Journal 82 (1977) 1013--1024.

\bibitem{Gingold:1977a}
R.~A. Gingold, J.~J. Monaghan, moothed particle hydrodynamics: theory and
  application to non-spherical stars, Mon. Not. Royal. Astr. Soc. 181 (1977)
  375--389.

\bibitem{Springel:2010/a}
V.~Springel, Smoothed particle hydrodynamics in astrophysics, Annual Review of
  Astronomy and Astrophysics 48~(1) (2010) 391--430.
\newblock \href {http://dx.doi.org/10.1146/annurev-astro-081309-130914}
  {\path{doi:10.1146/annurev-astro-081309-130914}}.

\bibitem{Gomez-Gesteira:2010/a}
M.~Gomez-Gesteira, B.~D. Rogers, R.~A. Dalrymple, A.~J. Crespo,
  State-of-the-art of classical sph for free-surface flows, Journal of
  Hydraulic Research 48 (2010) 6--27.
\newblock \href {http://dx.doi.org/10.1080/00221686.2010.9641242}
  {\path{doi:10.1080/00221686.2010.9641242}}.

\bibitem{Libersky:1991/a}
L.~D. {Libersky}, A.~G. {Petschek}, {Smooth particle hydrodynamics with
  strength of materials}, in: H.~E. {Trease}, M.~F. {Fritts}, W.~P. {Crowley}
  (Eds.), Advances in the Free-Lagrange Method Including Contributions on
  Adaptive Gridding and the Smooth Particle Hydrodynamics Method, Vol. 395 of
  Lecture Notes in Physics, Berlin Springer Verlag, 1991, pp. 248--257.
\newblock \href {http://dx.doi.org/10.1007/3-540-54960-9_58}
  {\path{doi:10.1007/3-540-54960-9_58}}.

\bibitem{Swegle:1995/a}
J.~W. Swegle, D.~L. Hicks, S.~W. Attaway, Smoothed particle hydrodynamics
  stability analysis, Journal of Computational Physics 116~(1) (1995) 123--134.

\bibitem{Dyka:1997/a}
C.~T. Dyka, P.~W. Randles, R.~P. Ingle, Stress points for tension instability
  in sph, International Journal for Numerical Methods in Engineering 40 (1997)
  2325--2341.
\newblock \href
  {http://dx.doi.org/10.1002/(SICI)1097-0207(19970715)40:13<2325::AID-NME161>3%
.0.CO;2-8}
  {\path{doi:10.1002/(SICI)1097-0207(19970715)40:13<2325::AID-NME161>3.0.CO;2-%
8}}.

\bibitem{Hicks:1997/a}
D.~L. Hicks, J.~W. Swegle, S.~W. Attaway, Conservative smoothing stabilizes
  discrete-numerical instabilities in {SPH} material dynamics computations,
  Applied Mathematics and Computation 85 (1997) 209--226.
\newblock \href
  {http://dx.doi.org/http://dx.doi.org/10.1016/S0096-3003(96)00136-1}
  {\path{doi:http://dx.doi.org/10.1016/S0096-3003(96)00136-1}}.

\bibitem{Monaghan:1989/a}
J.~J. Monaghan, On the problem of penetration in particle methods, J. Comput.
  Phys. 82 (1989) 1--15.
\newblock \href {http://dx.doi.org/10.1016/0021-9991(89)90032-6}
  {\path{doi:10.1016/0021-9991(89)90032-6}}.

\bibitem{Randles:1996a}
P.~Randles, L.~Libersky, {Smoothed Particle Hydrodynamics: Some recent
  improvements and applications}, Computer Methods in Applied Mechanics and
  Engineering 139~(1) (1996) 375--408.

\bibitem{Gray:2001/a}
J.~Gray, J.~Monaghan, R.~Swift, Sph elastic dynamics, Computer Methods in
  Applied Mechanics and Engineering 190 (2001) 6641 -- 6662.
\newblock \href
  {http://dx.doi.org/http://dx.doi.org/10.1016/S0045-7825(01)00254-7}
  {\path{doi:http://dx.doi.org/10.1016/S0045-7825(01)00254-7}}.

\bibitem{Belytschko:2000/a}
T.~Belytschko, Y.~Guo, W.~Kam~Liu, S.~Ping~Xiao, A unified stability analysis
  of meshless particle methods, International Journal for Numerical Methods in
  Engineering 48~(9) (2000) 1359--1400.

\bibitem{Bonet:2001/a}
J.~Bonet, S.~Kulasegaram, \href{http://dx.doi.org/10.1002/nme.242}{Remarks on
  tension instability of eulerian and lagrangian corrected smooth particle
  hydrodynamics (csph) methods}, International Journal for Numerical Methods in
  Engineering 52~(11) (2001) 1203--1220.
\newblock \href {http://dx.doi.org/10.1002/nme.242}
  {\path{doi:10.1002/nme.242}}.
\newline\urlprefix\url{http://dx.doi.org/10.1002/nme.242}

\bibitem{Bonet:2002/a}
J.~Bonet, S.~Kulasegaram, Alternative total lagrangian formulations for
  corrected smooth particle hydrodynamics ({CSPH}) methods in large strain
  dynamic problems, Revue Europ\'enne des \'El\'ements 11~(7-8) (2002)
  893--912.

\bibitem{Rabczuk:2004a}
T.~Rabczuk, T.~Belytschko, S.~Xiao, Stable particle methods based on lagrangian
  kernels, Computer Methods in Applied Mechanics and Engineering 193~(12-14)
  (2004) 1035--1063.

\bibitem{Vignjevic:2006a}
R.~Vignjevic, J.~Reveles, J.~Campbell, {SPH} in a total lagrangian formalism,
  Computer Modelling in Engineering and Sciences 146 (2)~(3) (2006) 181--198.

\bibitem{Xiao:2005/a}
S.~Xiao, T.~Belytschko, Material stability analysis of particle methods,
  Advances in Computational Mathematics 23~(1-2) (2005) 171--190.
\newblock \href {http://dx.doi.org/10.1007/s10444-004-1817-5}
  {\path{doi:10.1007/s10444-004-1817-5}}.

\bibitem{Flanagan:1981/a}
D.~P. Flanagan, T.~Belytschko, A uniform strain hexahedron and quadrilateral
  with orthogonal hourglass control, International Journal for Numerical
  Methods in Engineering 17 (1981) 679--706.
\newblock \href {http://dx.doi.org/10.1002/nme.1620170504}
  {\path{doi:10.1002/nme.1620170504}}.

\bibitem{Monaghan:1988/a}
J.~J. Monaghan, An introduction to {SPH}, Computer Physics Communications 48
  (1988) 89--96.
\newblock \href
  {http://dx.doi.org/http://dx.doi.org/10.1016/0010-4655(88)90026-4}
  {\path{doi:http://dx.doi.org/10.1016/0010-4655(88)90026-4}}.

\bibitem{Bonet:1999a}
J.~Bonet, T.-S. Lok, {Variational and momentum preservation aspects of Smooth
  Particle Hydrodynamic formulations}, Computer Methods in Applied Mechanics
  and Engineering 180~(1--2) (1999) 97--115.

\bibitem{Vignjevic:2000a}
R.~Vignjevic, J.~Campbell, L.~Libersky, {A treatment of zero-energy modes in
  the Smoothed Particle Hydrodynamics method}, Computer Methods in Applied
  Mechanics and Engineering 184~(1) (2000) 67--85.

\bibitem{Bonet:2008/a}
J.~Bonet, R.~D. Wood, Nonlinear Continuum Mechanics for Finite Element
  Analysis, Cambridge University Press, Cambridge, 2008.

\bibitem{Belytschko:2014/a}
T.~Belytschko, W.~K. Liu, B.~Moran, K.~Elkhodary, {Nonlinear Finite Elements
  for Continua and Structures, 2nd ed.}, John Wiley \& Sons, 2013.

\bibitem{Gingold:1982a}
R.~Gingold, J.~Monaghan, Kernel estimates as a basis for general particle
  methods in hydrodynamics, Journal of Computational Physics 46~(3) (1982)
  429--453.

\bibitem{Jameson:1985/a}
A.~Jameson, D.~Mavripilis, Finite volume solution of the two-dimensional euler
  equations on a regular triangular mesh, in: Proceedings of the AIAA 23rd
  Aerospace Sciences Meeting, AIAA paper 85-0435, 1985.

\bibitem{Desbrun:1996/a}
M.~Desbrun, M.~P. Cani, {Smoothed Particles: a new paradigm for animating
  highly deformable bodies}, in: Proceedings of Eurographics Workshop on
  Computer Animation and Simulation, Poitiers, France, 1996.

\bibitem{Timoshenko:1970/a}
S.~Timoshenko, J.~Goodier, Theory of Elasticity, McGraw-Hill, New York, 1970.

\bibitem{Chen:1996/a}
J.-S. Chen, C.~Pan, C.-T. Wu, W.~K. Liu, Reproducing kernel particle methods
  for large deformation analysis of non-linear structures, Computer Methods in
  Applied Mechanics and Engineering 139 (1996) 195--227.
\newblock \href {http://dx.doi.org/10.1016/S0045-7825(96)01083-3}
  {\path{doi:10.1016/S0045-7825(96)01083-3}}.

\bibitem{Vidal:2007/a}
Y.~Vidal, J.~Bonet, A.~Huerta, {Stabilized updated Lagrangian corrected {SPH}
  for explicit dynamic problems}, International Journal for Numerical Methods
  in Engineering 69~(13) (2007) 2687–2710.

\bibitem{Landau:1986/a}
L.~D. Landau, E.~M. Lifshitz, Theory of Elasticity, 3rd ed., Pergamon Press,
  1986.

\bibitem{Benz:1995/a}
W.~Benz, E.~Asphaug, Simulations of brittle solids using smooth particle
  hydrodynamics, Computer physics communications 87~(1) (1995) 253--265.

\bibitem{CesardeSa:2001/a}
J.~M.~A. César~de Sá, P.~M.~A. Areias, R.~M. Natal~Jorge, Quadrilateral
  elements for the solution of elasto-plastic finite strain problems,
  International Journal for Numerical Methods in Engineering 51~(8) (2001)
  883--917.
\newblock \href {http://dx.doi.org/10.1002/nme.183}
  {\path{doi:10.1002/nme.183}}.

\end{thebibliography}
\end{document}